\documentclass[a4paper,11pt]{article}
\usepackage{jcappub} 
\usepackage{lineno}

\usepackage{amssymb}
\usepackage{graphicx}
\usepackage{amsmath}

\usepackage{subfigure}
\usepackage{multirow}
\usepackage{setspace}
\usepackage{verbatim}
\usepackage{float}
\usepackage{color}

\usepackage[utf8]{inputenc}
\usepackage[table,xcdraw]{xcolor}

\newcommand{\md}{\mathrm{d}}

\arxivnumber{2312.14041} 

\title{\boldmath Distinctive GWBs from eccentric inspiraling SMBH binaries with a DM spike}

\author[a,b,c]{Li Hu}
\author[d,b,a]{Rong-Gen Cai}
\author[b,e,1]{Shao-Jiang Wang\note{Corresponding author.}}
\affiliation[a]{School of Fundamental Physics and Mathematical Sciences, Hangzhou Institute for Advanced Study (HIAS), University of Chinese Academy of Sciences (UCAS),\\Hangzhou 310024, China}
\affiliation[b]{CAS Key Laboratory of Theoretical Physics, Institute of Theoretical Physics, Chinese Academy of Sciences (CAS),\\Beijing 100190, China}
\affiliation[c]{University of Chinese Academy of Sciences (UCAS),\\Beijing 100049, China}
\affiliation[d]{School of Physical Science and Technology, Ningbo University,\\Ningbo 315211, China}
\affiliation[e]{Asia Pacific Center for Theoretical Physics (APCTP),\\Pohang 37673, Korea}
\emailAdd{schwang@itp.ac.cn}

\abstract{
Recent detections of a low-frequency gravitational wave background (GWB) from various pulsar-timing-array (PTA) observations have renewed the interest in the inspiraling supermassive black hole binaries (SMBHBs), whose population is believed to be the most promising candidate with possible generalizations from including either orbital eccentricity or dark matter (DM) spike. In this paper, we show that the inclusion of both can further display distinctive features detectable in future PTA observations. With a typical initial eccentricity $e_0\sim\mathcal{O}(0.1)$ for the inspiraling SMBHBs, even a shallow DM spike can easily drive the orbital eccentricity close to $1$, leaving behind a large turnover eccentricity when GWs begin to dominate the orbital circularization. In particular, the DM spike index $\gamma_\mathrm{sp}$ universally manifests itself in the characteristic strain by $h_c\sim f^{7/6-\gamma_\mathrm{sp}/3}$ in the far infrared and features a novel oscillation structure at low frequencies. Future PTA detection of such characteristics would be the smoking gun for the DM spike and even reveal the nature of DM.}

\begin{document}
\maketitle
\flushbottom


\section{Introduction} 
One-fifth content of our current Universe is believed to be contributed by dark matter (DM), whose very existence is indirectly supported by astrophysical and cosmological observations over various ranges of scales, including the rotation curves~\cite{Rubin:1970zza,Begeman:1991iy}, bullet clusters~\cite{Zwicky:1933gu,Clowe:2006eq}, large scale structures~\cite{Dietrich:2012mp}, and cosmic microwave background anisotropics~\cite{Planck:2018vyg}. However, both direct and indirect DM detections return null results~\cite{PandaX:2022osq,XENON:2023cxc,DiMauro:2023qat}, calling for more pronounced probes~\cite{Baryakhtar:2022hbu} into the nature of DM. One such a promising target is at the center of the galaxy usually hosting a supermassive black hole (SMBH)~\cite{Richstone:1998ky,Kormendy:2013dxa} of mass from $10^{6}\,M_{\odot}$ to $10^{10}\,M_{\odot}$, which can steepen the DM density profile at galactic center into a dense spike~\cite{Gondolo:1999ef}, largely enhancing the prospects for indirect detection, in particular for a rotating SMBH~\cite{Ferrer:2017xwm}. Nevertheless, the galactic foregrounds might obscure the annihilation signals from the DM spike, and hence one can turn to the gravitational-wave (GW) probe that is transparent to galactic intermedium as an alternative examination of DM properties.

One way to generate GWs from SMBHs is through the merger of two galaxies~\cite{Lotz:2011cn} frequently in hierarchical systems~\cite{Lacey:1993iv} with their dual SMBHs spiraling in and settling to~\cite{DeRosa:2019myq} the new center of the merged galaxy~\cite{Barnes:1992rm} by dynamical friction~\cite{Chandrasekhar:1943ys,Antonini:2011tu}. After that, although the dynamical friction becomes less efficient, there can be a stage when dual SMBHs are still quite far apart but eventually bounded by gravity as a supermassive black hole binary (SMBHB)~\cite{Begelman:1980vb,Merritt:2004gc} at parsec-scale separation. At this stage, however, the pure GW-driven~\cite{Peters:1964zz} inspiral of a SMBHB produces GW frequencies way below the pulsar-timing-array (PTA) band, and cannot coalesce within the Hubble time, leading to the long-standing final-parsec problem~\cite{Begelman:1980vb} that necessitates binary hardenings~\cite{Yu:2001xp,Sesana:2013wja} from stellar and gaseous environment~\cite{2012AdAst2012E...3D} around the SMBHB. Here we simply assume that the final-parsec problem has been overcome by the dynamical friction from the DM spike of self-interacting type as suggested by a recent study~\cite{Alonso-Alvarez:2024gdz}~\footnote{It follows that $t_\mathrm{df}\lesssim 1$ Gyr as long as the DM spike index $\gamma_\mathrm{sp}\gtrsim 0.7$ as we will set in the following study.}. When the final-parsec problem can be overcome, GWs from the inspiral phase of an SMBHB are the most promising GW sources in the PTA band~\cite{NANOGrav:2019tvo}, and hence the following GWs from the coalescence phase of an SMBHB can be expected in future space-borne GW detectors~\cite{Bi:2023tib}. See also Ref.~\cite{Daghigh:2022pcr} for GWs from the ringdown phase of galactic BHs surrounded by a DM spike.

Although currently no continuous signal of any individual SMBHB merger has been reported so far, recent PTA observations from the North American Nanohertz Observatory for Gravitational Waves (NANOGrav)~\cite{NANOGrav:2023gor}, European Pulsar Timing Array (EPTA)~\cite{EPTA:2023sfo}, Parkes Pulsar Timing Array (PPTA)~\cite{Reardon:2023gzh}, and Chinese Pulsar Timing Array (CPTA)~\cite{Xu:2023wog} have found independent evidence for detecting a stochastic GW background (SGWB) around the nano-Hertz band. In particular, the NANOGrav 15yr data prefers the environmentally driven over the GW-only driven binary evolutions, especially to the lowest frequency
bin~\cite{NANOGrav:2023hfp}, where the GW spectrum of the characteristic strain $h_c$ tends to deviate slightly at low frequencies from the well-known power-law $h_c\propto f^{-2/3}$~\cite{Maggiore:2018sht}. Including either the eccentricity~\cite{Bi:2023tib} or environmental effects~\cite{Ellis:2023dgf}, for example, DM profiles~\cite{Ghoshal:2023fhh}, DM spike~\cite{Shen:2023pan},  and DM soliton~\cite{Aghaie:2023lan}, all lead to changes in the power spectrum profile, and some of them may provide more accurate templates for future GW detections.

Nevertheless, a word of caution should be taken to motivate any further interpretations on top of what is constrained by circular GW-driven binaries from a few low-frequency bins of current PTA data. This is because the spectrum is still extremely uncertain and the $h_c\propto f^{-2/3}$ power law is not necessarily the expectation from circular GW-driven binaries due to the discreteness effects~\cite{EPTA:2023xxk}. The discreteness effects come into play in the following two aspects~\cite{Agazie:2024jbf}: the excursions from a few nearby and/or very massive SMBHBs before being individually resolvable, and the breaks in the stochasticity of SGWBs at some frequency bins with insufficient sources. As a result, there is actually no need to claim either eccentricity or environmental effects to interpret recent PTA observations. This does not mean that they do not play a role or eventually emerge in future PTA observations, but only the current investigation is of purely theoretical interest.

It is worth noting that even a moderate DM spike is known to boost the eccentricity up to a near-extreme value for intermediate mass ratio inspirals (IMRIs)~\cite{Yue:2019ozq}, making us wonder if it is also the case for SMBHB inspirals, which, to our best knowledge, has not been fully explored before when both eccentricity and DM spike are considered. Note that the presence of the DM spike is not necessary for addressing the last-parsec problem that has already been solved by the usual stellar hardening. Here we only mean to investigate in detail the impact from the DM spike index $\gamma_\mathrm{sp}$ on the characteristic strain $h_c$ of inspiraling SMBHBs when there is an initial eccentricity, and reveal a universal scaling law $h_c\sim f^{7/6-\gamma_\mathrm{sp}/3}$ for the spectrum in the far infrared regime. Moreover, on top of the turnover-bump-flatter spectrum with orbital eccentricity~\cite{Enoki:2006kj,Sesana:2013wja,Huerta:2015pva,Chen:2016zyo,Kelley:2017lek}, the inclusion of the DM spike could further generate an oscillation feature at low frequencies, signifying distinctive perspective of DM probe in future PTA observations.

\section{Eccentric inspiral in the DM spike}

Consider a SMBHB consisting of two SMBHs with masses $m_1$ and $m_2$ moving along elliptical orbits of a common eccentricity $e$ with semi-major axes $a_1$ and $a_2$ and distances $r_1$ and $r_2$ to the barycenter satisfying $m_1/m_2=r_2/r_1=a_2/a_1$, respectively, the secular dynamics of each BH, for example, the BH $m_1$, is equivalent to first remove the other BH $m_2$ and then place an object of mass $M_1=m_2^3/M^2$ at the barycenter where $M\equiv m_1+m_2$, and hence the secular dynamics of $m_1$ simply follows~\cite{Yue:2019ozq}
\begin{align}
r_1&=\frac{a_1(1-e^2)}{1+e\cos{\phi_1}},\\
E_1&=-\frac{Gm_1M_1}{2a_1},\\
L_1&=m_1\sqrt{M_1Ga_1(1-e^2)},
\end{align}
where $\phi_1$, $E_1$, and $L_1$ are the angular position, energy, and angular momentum of the BH $m_1$, respectively, and $G$ is the Newtonian constant. Therefore, after introducing $a\equiv a_1+a_2$ as the semi-major axis of the whole orbit, the total energy $E$ and total angular momentum $L$ of SMBHB are~\cite{Peters:1964zz}
\begin{align}
E&=E_1+E_2=-\frac{Gm_1m_2}{2a},\label{eq:E}\\
L&=L_1+L_2=\frac{m_1m_2\sqrt{Ga(1-e^2)}}{\sqrt{M}}.\label{eq:L}
\end{align}

Now we further include the DM spike. Given an initial DM halo profile $\rho_\mathrm{halo}(r)$ developing an inner cusp $\rho_\mathrm{halo}(r\ll r_0)=\rho_0(r/r_0)^{-\gamma_0}$ below some characteristic scale $r_0$, the adiabatic growth of the DM profile around the galactic BH leads to the formation of a DM spike~\cite{Gondolo:1999ef},
\begin{align}
\rho_\mathrm{spike}(r_\mathrm{ISCO}<r<R_\mathrm{sp})=\xi\rho_\mathrm{sp}\left(\frac{R_\mathrm{sp}}{r}\right)^{\gamma_\mathrm{sp}},
\end{align}
where the original inner slope $\gamma_0$ within a typical range $0\leq\gamma_0\leq2$ is steepened into $\gamma_\mathrm{sp}\equiv(9-2\gamma_0)/(4-\gamma_0)$ within the new range $2.25\leq\gamma_\mathrm{sp}\leq2.5$. This DM spike profile should be cut off below the innermost stable circular orbit (ISCO) of the central BH, and is extended up to a spike radius scale $R_\mathrm{sp}\propto r_0(M_\mathrm{BH}/\rho_0r_0^3)^{1/(3-\gamma_0)}$. For more realistic modelings of environmental effects of SMBH~\cite{Ostriker:1999ee,Ullio:2001fb,Merritt:2002vj,Gnedin:2003rj,Merritt:2003pf,Merritt:2003qk,Shapiro:2022prq}, the DM spike slope does not need to follow the adiabatic growth $\gamma_\mathrm{sp}\equiv(9-2\gamma_0)/(4-\gamma_0)$, and hereafter we can treat $\gamma_\mathrm{sp}$ generally as a free parameter. The factor $\xi$ represents the suppression of the self-interacting dark matter spike relative to the cold dark matter spike. The determinations for the other two parameters $\rho_\mathrm{sp}$ and $R_\mathrm{sp}$ in the DM spike profile go parallel to that of Ref.~\cite{Shen:2023pan}, which can be traced back to the central BH mass $M_\mathrm{BH}$ at a redshift $z$ alone as illustrated in details in Appendix~\ref{Spike} for a Navarro-Frenk-White (NFW) DM halo profile $\rho_\mathrm{NFW}(r)=\rho_s(r/r_s)^{-1}(1+r/r_s)^{-2}$~\cite{Navarro:1996gj} with $\rho_0\equiv\rho_s$, $r_0\equiv r_s$, and $\gamma_0=1$.

\begin{figure}[htbp]
\centering
\includegraphics[width=0.75\textwidth]{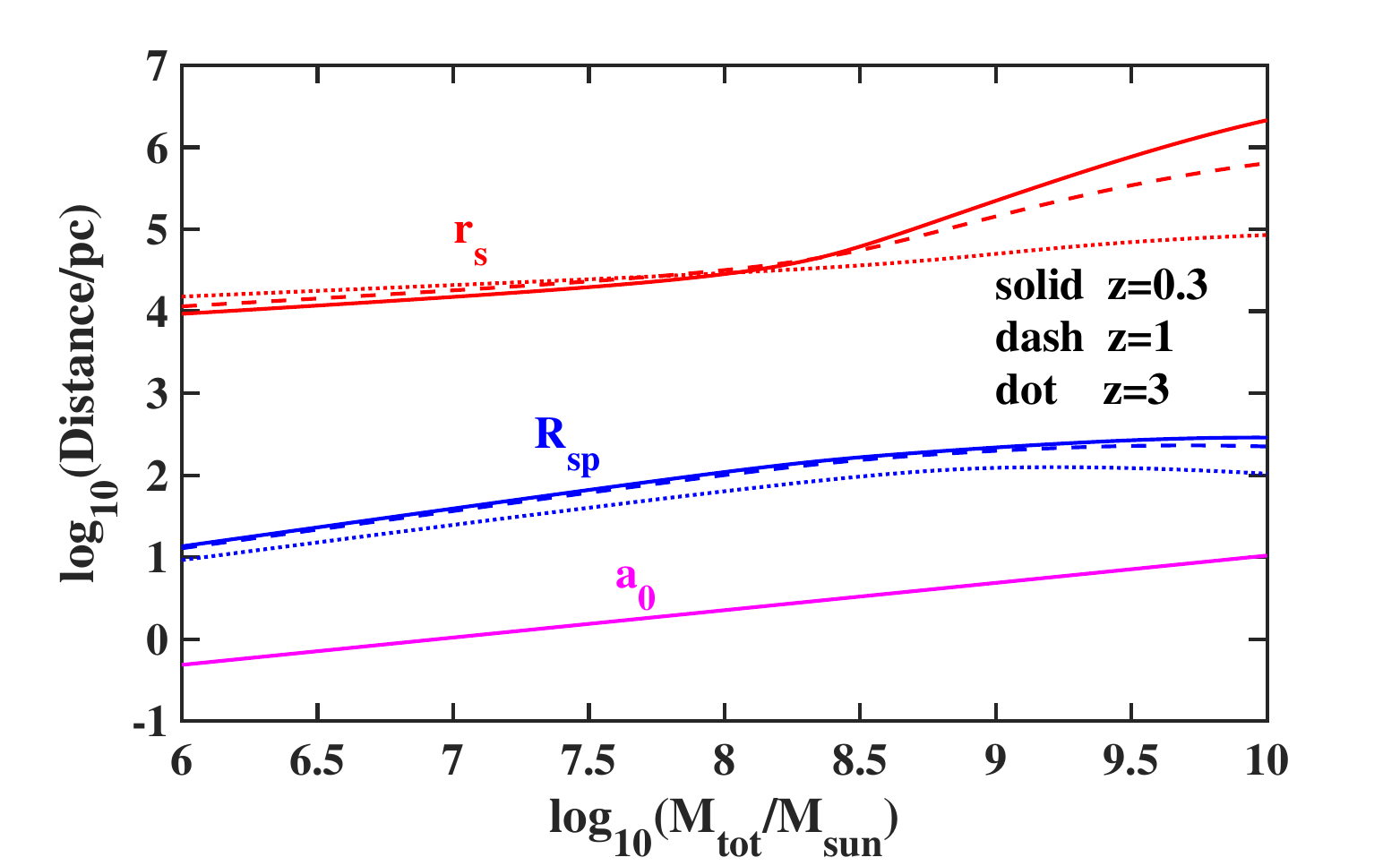}
\caption{Comparison of $r_s$, $R_\mathrm{sp}$, and $a_0$ of the SMBHB with respect to the total BH mass at different redshifts.}
\label{fig:distance}
\end{figure}

Usually, the scale radius $r_s$ is roughly three orders of magnitude larger than $R_\mathrm{sp}$, and $R_\mathrm{sp}$ is nearly two orders of magnitude larger than the initial semi-major axis $a_0$ we set in this study as shown in Fig.~\ref{fig:distance}, therefore, we can safely approximate the total DM spike of the SMBHB as the one in the barycenter effectively induced by the total BH mass $M=m_1+m_2$. When each BH travels through this DM spike, it will simultaneously experience a well-known dynamical friction force~\cite{Chandrasekhar:1943ys,Binney:2008} (See also Ref.~\cite{Dosopoulou:2023umg} for improved treatments)
\begin{align}
F_i=\frac{4\pi G^2m_i^2\rho_\mathrm{spike}(r_i)\ln\Lambda}{v_i^2}, \quad i=1,2
\end{align}
that goes against its moving direction due to the gravitational interactions with DM particles, where $v_i$ is the velocity of BH $i$ in the center-of-mass system of SMBHB, $\ln\Lambda$ is the Coulomb logarithm usually chosen as $\ln\Lambda=3$ in the literature~\cite{Burke-Spolaor:2018bvk}. The loss rates of energy and angular momentum from the dynamical friction and the evolution of the orbital semi-major axis $a$ and eccentricity $e$ are calculated in detail in Appendix~\ref{EOM}.

Here we provide a summary of the main contents of Appendix~\ref{EOM}. The loss rate caused by GWs has already been obtained to the lowest order in post-Newtonian expansion~\cite{Peters:1964zz}. Meanwhile, the average loss rate due to dynamical friction can be calculated directly through Newtonian mechanics with respect to the orbital period. Then, the energy and angular momentum loss rates from both GWs and dynamical friction can be inserted into Eqs. \eqref{eq:E}, \eqref{eq:L} to obtain the dynamical equations governing the evolution of the orbital semi-major axis $a$ and eccentricity $e$.

\section{GW backgrounds from SMBHBs}
Numerically solving for $a(t)$ and $e(t)$ from their dynamical equations, we can obtain two GW polarizations as shown in Appendix~\ref{GW} to be further expanded as~\cite{Moore:2018kvz}
\begin{align}\label{eq:series}
h_{+,\times}=A\sum_{n=1}^{\infty}\left(C_{+,\times}^{(n)}\cos(n\ell)+S_{+,\times}^{(n)}\sin(n\ell)\right),
\end{align}
where coefficients $C_{+,\times}^{(n)}$ and $S_{+,\times}^{(n)}$~\cite{Moore:2018kvz,Becker:2021ivq} are attached in Appendix~\ref{GW}. Both the GW amplitude $A=-\frac{G^{5/3}M_{c}}{c^4D_{L}}(2\pi M_{c}F)^{2/3}$ and the mean anomaly $\ell(t)=\int^{t}\mathrm{d}t\,2\pi F$ are computed from the mean orbital frequency $F=\sqrt{GM/a^3}/(2\pi)$, where $M_c=\mu^{3/5}M^{2/5}$ is the chirp mass, $\mu=m_1m_2/M$ is the reduced mass,  and the luminosity distance $D_L$ will be eventually canceled out in the energy spectrum. As usual, the stationary phase approximation is adopted for Fourier modes of two GW polarizations since the amplitude $A$ varies slowly with time, in which case only those GW frequencies satisfying stationary phase condition $f=nF(t_n^{*})$ can contribute non-negligibly to the Fourier integral, and hence the orbital frequency $F$ no longer holds the one-to-one correspondence to the GW frequency $f$ as the circular binary. Here $t_n^\ast$ is the “stationary” time for the $n^{th}$ harmonic which relates the
Fourier frequency $(f)$ to the mean orbital frequency $(F)$ through the stationary phase condition. Therefore, the Fourier-transformed GW polarizations are computed by
\begin{align}
\left|\tilde{h}_{+,\times}^{(n)}(f)\right|
=\frac{G^{5/3}M_{c}}{2c^4D_{L}}\frac{(2\pi M_cF(t_n^{*}))^{2/3}}{\sqrt{n\dot{F}(t_n^{*})}}\times\left|{C}_{+,\times}^{(n)}(t_n^{*})+\dot{\imath}S_{+,\times}^{(n)}(t_n^{*})\right|\,.
\end{align}

The inclusion of the eccentricity would have to sum over all integer harmonics especially $n>2$ but practically truncated at, for example, $n=1000$ as we have checked numerically for the convergence of $\tilde{h}_{+,\times}(f)=\sum_{n=1}^{1000}\tilde{h}_{+,\times}^{(n)}(f)$. Under the mutually orthogonal approximation between different harmonics with controlled error~\cite{Moore:2018kvz}, the GW energy spectrum can be approximately calculated as~\cite{Maggiore:2007ulw}
\begin{align}
\frac{\mathrm{d}E_\mathrm{GW}}{\mathrm{d}f_r}
=\frac{\pi c^3}{2G}\frac{f^2D_L^2}{(1+z)^2}\int \mathrm{d}\Omega\left(\left|\tilde{h}_{+}(f)\right|^2+\left|\tilde{h}_{\times}(f)\right|^2\right),
\end{align}
where $\rm d \Omega$ is the solid angle element, which can be decomposed into $\sin{(\iota)}\rm{d}\iota\rm{d}\beta$, furthermore, the definition of $\iota$ and $\beta$ can be found in Appendix~\ref{GW}. Then
the characteristic strain $h_c$ is obtained by integrating the above GW energy spectrum  over all different SMBHB populations as
\begin{align}
h_c^2(f)=\frac{4G}{\pi fc^2}\int\int\int \mathrm{d}z\mathrm{d}m_1\mathrm{d}q\frac{\mathrm{d}^3n}{\mathrm{d}z\mathrm{d}m_1\mathrm{d}q}\frac{\mathrm{d}E_\mathrm{GW}}{\mathrm{d}f_r}.
\end{align}
Here each SMBHB is characterized by the mass ratio $q=m_2/m_1$ with respect to one of the BH mass $m_1$ at redshift $z$, and $f_r=f(1+z)$ is the frequency measured at the source rest frame. The integration ranges are chosen loosely as $0\leq z\leq5$, $10^{6}\leq M_1/M_{\odot}\leq10^{10}$, $10^{-3}\leq q\leq1$ for a general consideration. The SMBHB population is parameterized by the same model used in Ref.~\cite{Chen:2018znx} as
\begin{align}
\frac{\mathrm{d}^3n}{\mathrm{d}z\mathrm{d}m_1\mathrm{d}q}
=\dot{n}_0\left[\left(\frac{m_1}{10^{7}\,M_{\odot}}\right)^{-\alpha}e^{-m_1/M_*}e^{-z/z_0}\right]\times(1+z)^{\beta_z}q^{\gamma}\frac{\mathrm{d}t_r}{\mathrm{d}z}\,, \quad \frac{\mathrm{d}t_r}{\mathrm{d}z}=\frac{-1}{(1+z)H(z)}\,,
\end{align}
with the same parameter set $(\lg\dot{n}_0,\lg M_*,z_0,\alpha,\beta_z,\gamma)=(-38.2,7.86,0.23,-1.35,-1.63,0.4)$ fixed as the one obtained in Ref.~\cite{Shen:2023pan} since we only care about the DM spike effect on the eccentric inspiral in this study. Note that we illustrate conservatively with $\gamma_\mathrm{sp}=0.5$ in Figs.~\ref{fig:Res_fixgamma}, \ref{fig:Res_WithoutDM}, and~\ref{fig:Res_varybase} below since $\gamma_\mathrm{sp}\gtrsim0.92$ has been ruled out at a $95\%$ confidence level from stellar orbits around Sagittarius A*~\cite{Shen:2023kkm}. In fact, we have also checked explicitly that the enclosed mass within the S2 orbit ($\sim 1$ mpc) of Sagittarius A* is much less than $10^3\,M_\odot$ ($\sim0.1\%$ of the SMBH mass) for $0\lesssim\gamma_\mathrm{sp}\lesssim1.5$. The full investigation of all parameters is preserved in the future.

\section{Results and analysis}

\begin{figure}[htbp]
\centering
\includegraphics[width=0.74\textwidth]{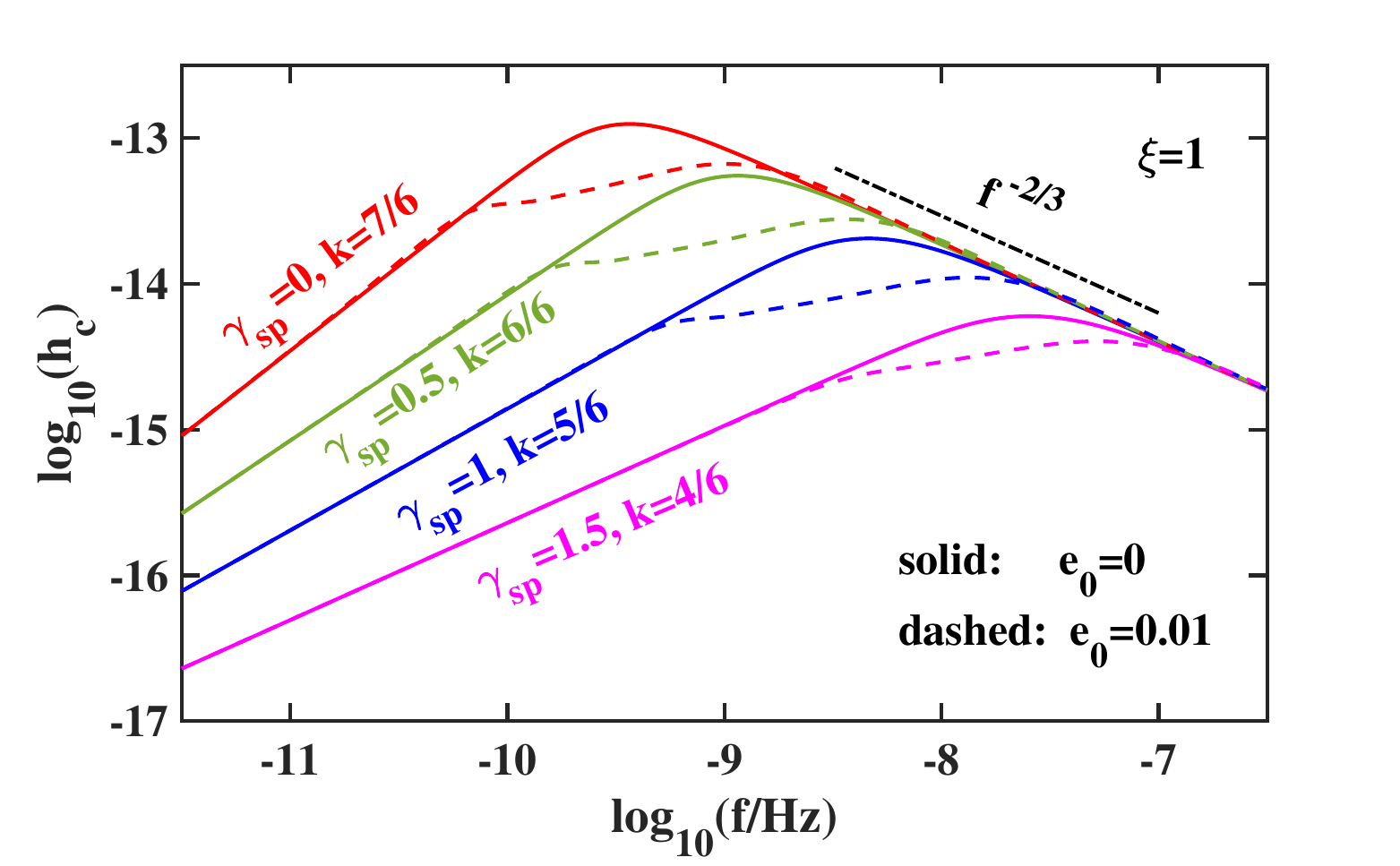}
\caption{The characteristic strain $h_c(f)$ at each GW frequency $f$ for the SMBHB inspiral with negligible initial eccentricity $e_0$ but also with different DM spike index $\gamma_\mathrm{sp}$. $k$ is the fitted infrared slope at low frequencies when DM spike is dominated, and $\xi=1$ is the suppression factor.}
\label{fig:Res_fixe0}
\end{figure}

We start our eccentric inspirals of SMBHBs in the DM spike at the same initial semi-major axis $a_0$ determined by the mean orbital frequency $F_0=10^{-13}$ Hz. Let us first look into the DM spike effect for a negligible initial eccentricity as shown in Fig.~\ref{fig:Res_fixe0}. Compared to the single power law for GW-only circular evolution, the eccentric binaries can emit GWs at all integer harmonics, boosting GW emissions from lower frequencies to higher frequencies. For an initial circular orbit, $e_0=0$, a constant DM profile without a spike, $\gamma_\mathrm{sp}=0$, would eventually flatten the spectrum and then turn over the single power law $h_c\sim f^{-2/3}$ into a broken power law with $h_c\sim f^{7/6}$ at low frequencies as seen from the red solid curve. The extra inclusion of DM spike with an increasing index $\gamma_\mathrm{sp}$ would further widen the infrared part of the broken power law at low frequencies with a decreasing slope,
\begin{align}
h_c\sim f^{7/6-\gamma_\mathrm{sp}/3}.
\end{align}

\begin{figure}[htbp]
\centering
\includegraphics[width=0.74\textwidth]{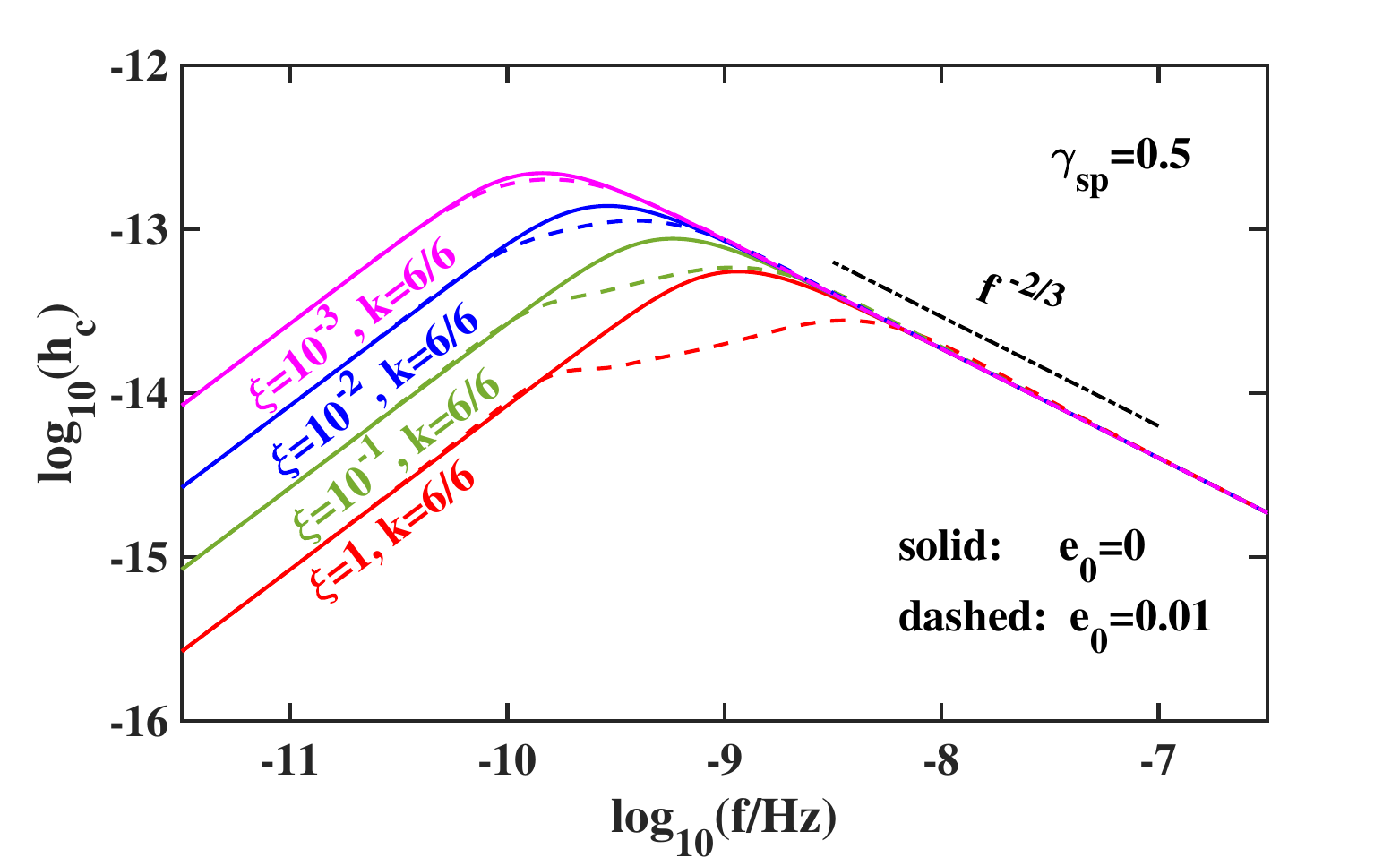}
\caption{The characteristic strain $h_c(f)$ at each GW frequency $f$ for the SMBHB inspiral with negligible initial eccentricity $e_0$ but also with different suppression factor $\xi$. $k$ is the fitted infrared slope at low frequencies when DM spike is dominated, and $\gamma_\mathrm{sp}=0.5$ is the spike index.}
\label{fig:xi_vary}
\end{figure}

For a marginally small initial eccentricity, for example, $e_0=10^{-2}$, the peak of the broken power law would be further flattened into an intermediate power-law section for shallower DM spikes, and the frequency ranges dominated by DM and GW will shrink accordingly. Although the initial eccentricity $e_0$ is marginally small, it can still grow to a considerable value in the process of evolution, and cause a certain degree of depression in $h_c$. This relatively flat transition section is similar to the deviation from the single power law in the NANOGrav 15-yr data with the help of DM spike for circular binaries~\cite{Shen:2023pan}. Only in our case, we do not need such a large DM spike index anymore~\cite {Shen:2023kkm}, but a marginally small initial eccentricity in a shallower DM spike would do the same job. 

\begin{figure}[htbp]
\centering
\includegraphics[width=0.74\textwidth]{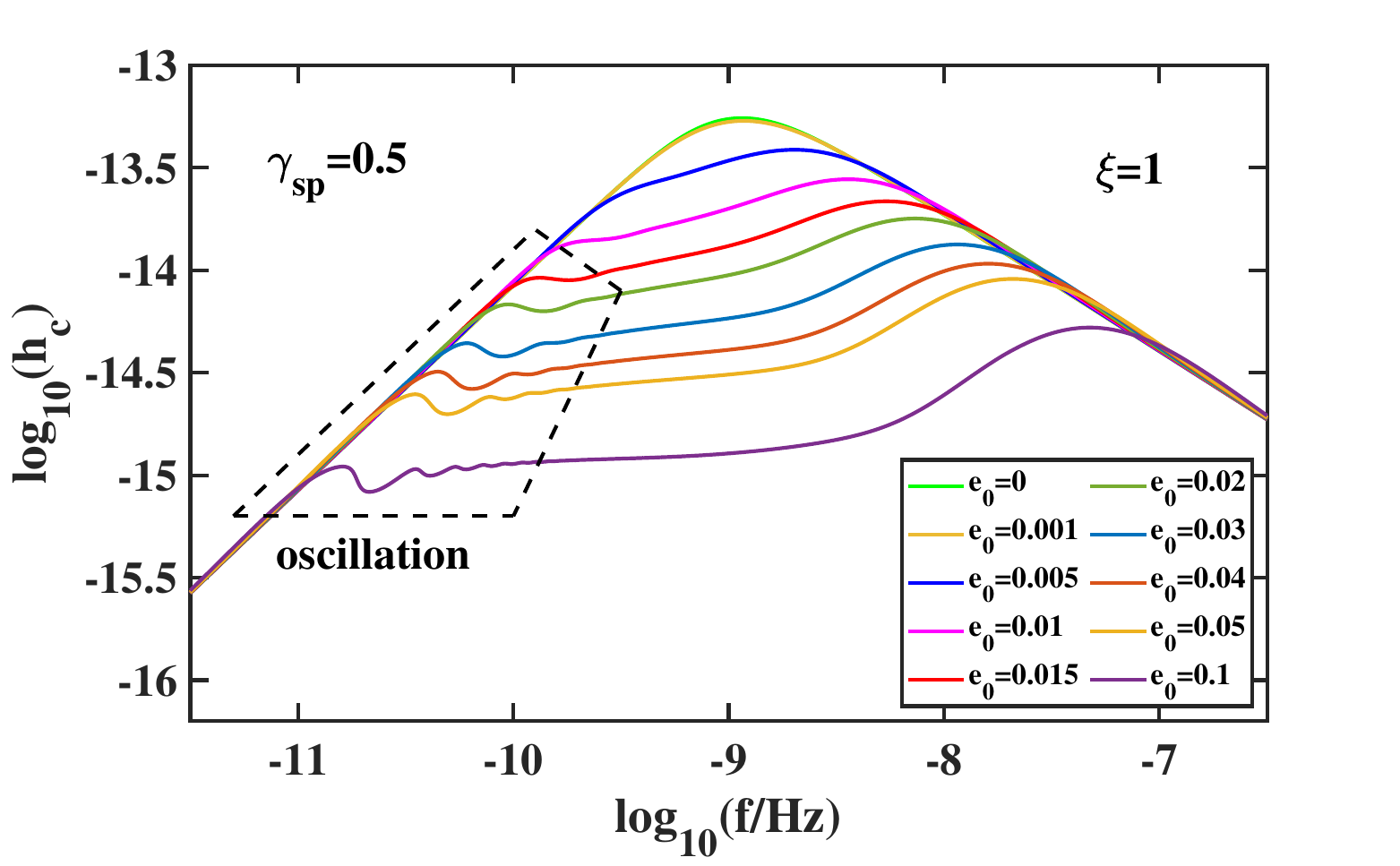}
\caption{The characteristic strain of GWs from SMBHBs with a fixed DM spike index $\gamma_\mathrm{sp}=0.5$ and suppression factor $\xi=1$, but for the initial eccentricity $e_0$ increasing from $0$ to $10^{-1}$.}
\label{fig:Res_fixgamma}
\end{figure}

\begin{figure}[hbp]
\centering
\includegraphics[width=0.75\textwidth]{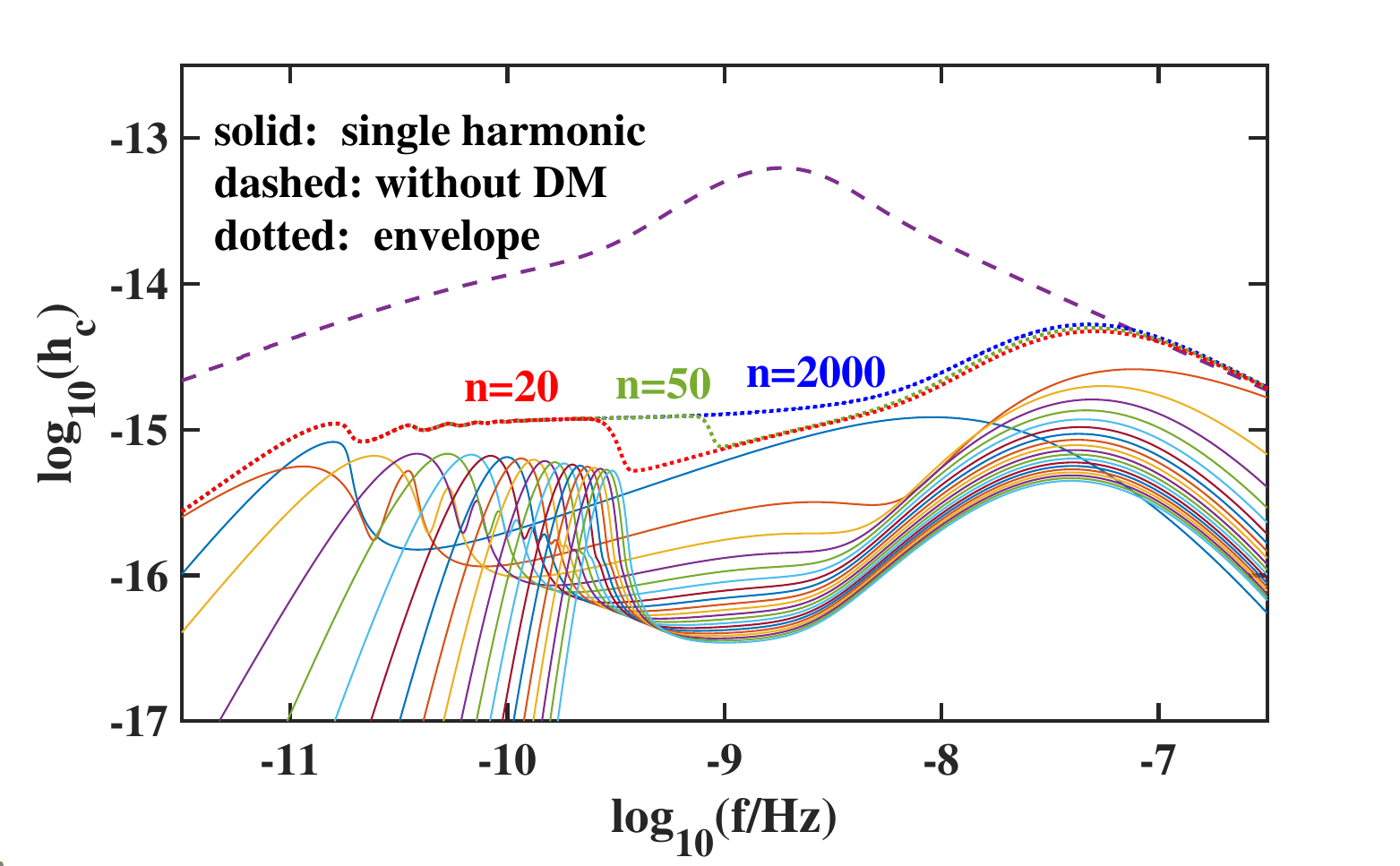}
\caption{The characteristic strain $h_c(f)$ of GWs with different initial settings. The dashed curve comes with a near-extreme initial eccentricity $e_0=0.99$ but without a DM environment. The dotted curves are for a fixed DM spike index $\gamma_\mathrm{sp}=0.5$, initial eccentricity $e_0=0.1$, and suppression factor $\xi=1$, but with the number of harmonics increasing from $n=20$ (red dotted) to $n=50$ (green dotted) and eventually converging to the case of $n=2000$ (blue dotted). The solid curves show $h_c(f)$ in the $n$-th harmonics with $n$ ranging from $1$ to $20$.}
\label{fig:Res_WithoutDM}
\end{figure}

Considering a range of possible suppression factors $\xi$ that represent the suppression of the self-interacting dark matter spike
relative to the cold dark matter spike as shown in Fig.~\ref{fig:xi_vary}. It can be seen that as the suppression factor decreases (namely an increasing suppression effect), the peak of the broken power law would move to the upper left of the figure, which is similar to the effect of reducing $\gamma_\mathrm{sp}$. This suggests that compared to cold dark matter, self-interacting dark matter tends to have a larger $\gamma_\mathrm{sp}$ to explain the PTA data. Meanwhile, as the suppression factor decreases, one significant difference compared to the reducing of $\gamma_\mathrm{sp}$ is that the fitted infrared slope $k$ at low frequencies remains a constant since $\gamma_\mathrm{sp}$ is fixed, and the other is the difference between two $h_c$ with different initial eccentricities would decrease (since the decrease of dark matter density).

\begin{figure}
\centering
\includegraphics[width=0.75\textwidth]{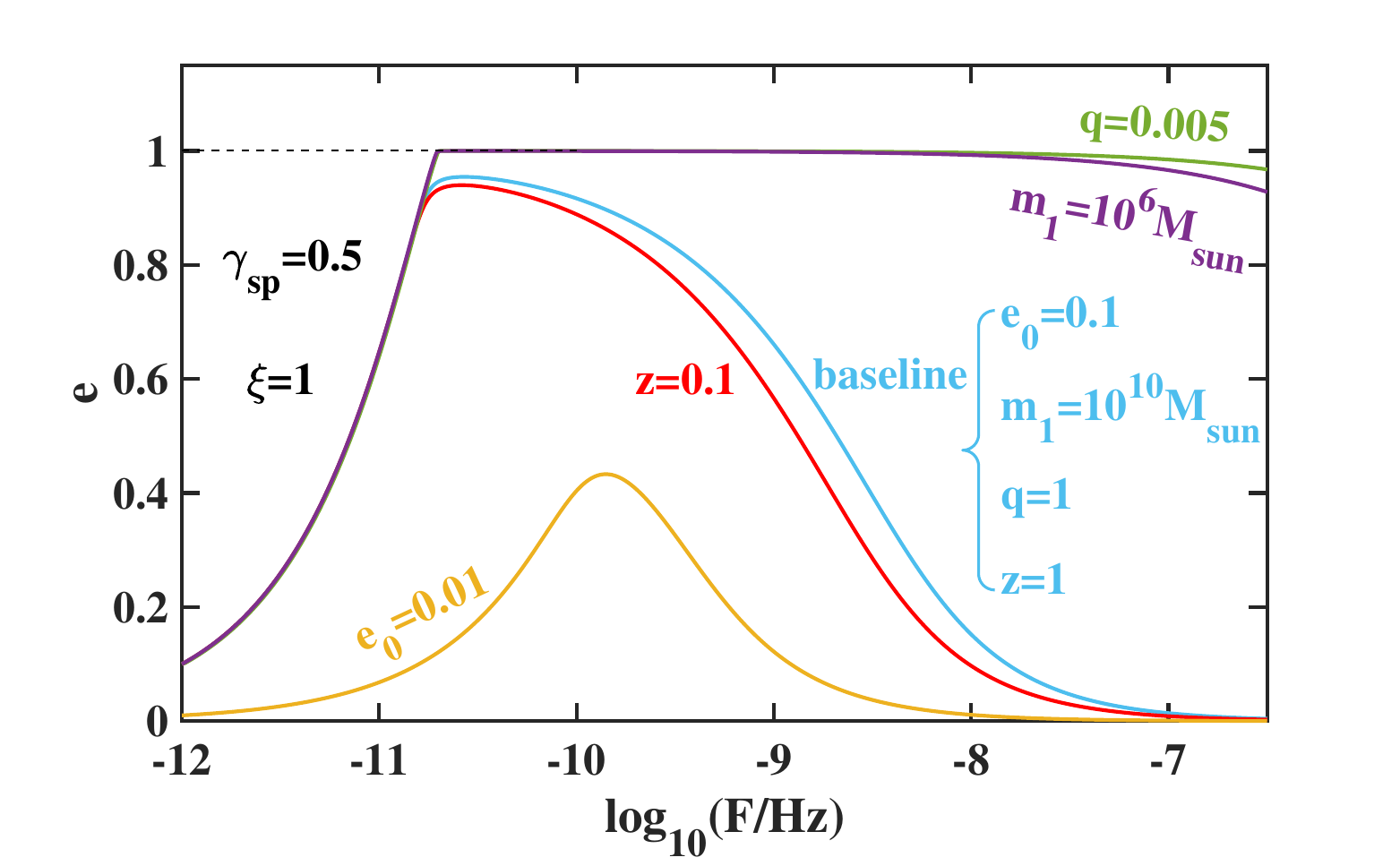}
\caption{Dynamical evolution of orbital eccentricity $e$ with orbital frequency $F$ for the baseline parameters $e_0=0.1$, $m_1=10^{10}\,M_{\odot}$, $q=1$, and $z=1$ along with its deviations in each one of the baseline parameters.}
\label{fig:Res_varybase}
\end{figure}

For the initial eccentricity increasing from $e_0=0$ to a marginally small value $e_0=10^{-2}$ and finally to an ordinarily small value $e_0=10^{-1}$ as shown in Fig.~\ref{fig:Res_fixgamma} with $\gamma_\mathrm{sp}=0.5$ fixed by considering the backreaction from the SMBHB formation to the DM spike~\cite{Merritt:2002vj}, the usual turnover-bump-flatter GW spectrum~\cite{Enoki:2006kj,Sesana:2013wja,Huerta:2015pva,Chen:2016zyo,Kelley:2017lek,NANOGrav:2023hfp} in the characteristic strain $h_c(f)$ gradually exhibits a new non-trivial structure with more and more oscillations at lower and lower frequencies. These oscillations can be viewed as the envelope of the superposition of different harmonics as shown in Fig.~\ref{fig:Res_WithoutDM}, where we illustrate with a red dotted curve for the superposition of the first 20 harmonics shown in solid colored curves with $e_0=0.1$, $\gamma_\mathrm{sp}=0.5$ and $\xi=1$. To better understand these oscillations, one can choose an arbitrary single harmonic with $2\leq n\leq20$ and focus on the low-frequency band corresponding to the oscillation feature, where a lower peak always follows a higher peak. The underlying reason behind this is that the contribution of the coefficients $C_{+,\times}^{(n)}$ and $S_{+,\times}^{(n)}$ in Eq.~(\ref{eq:series}) to a single harmonic for any given $n$, $C_{+}^2+C_{\times}^{2}+S_{+}^{2}+S_{\times}^{2}$, is not a monotonic function of the eccentricity $e$. In fact, as $e$ increases, the total contribution drops rapidly after reaching a peak (the position of the peak roughly corresponds to $e=0.8$, which depends on $n$), which would cause $h_c$ of single harmonic to be significantly suppressed when the eccentricity is large. Note that in this case, the contribution of the baseline in Fig.~\ref{fig:Res_varybase} dominates, and its eccentricity corresponds to an increase, decrease, and then increase in $C_{+}^2+C_{\times}^{2}+S_{+}^{2}+S_{\times}^{2}$, resulting in two peaks of single harmonic $h_c$. As for the height difference between the two peaks, the contribution of $a$ and $\dot a$ should also be considered.

When summing more harmonics, this oscillation feature remains intact but only with the dip moving to a higher frequency until it totally disappears, as shown illustratively with green and blue dotted curves from summing $n=50, 2000$ harmonics, respectively. Note here that we have checked numerically the results will converge for $n\gtrsim1000$ as the total signal from summing $n=1000$ harmonics is almost indistinguishable from that of the $n=2000$ case.

As a comparison, in the absence of the DM spike but with a near-extreme initial eccentricity $e_0=0.99$, the characteristic strain admits no oscillations as shown in Fig.~\ref{fig:Res_WithoutDM} with a dashed curve, this result is similar to Fig.1. of \cite{Chen:2016zyo}. Therefore, future PTA detection of this oscillation feature would necessarily provide the smoking gun for the DM spike or even reveal the nature of the DM. Note here that the near-extreme initial eccentricity $e_0=0.99$ in the absence of the DM spike is chosen for comparison to the case with a DM spike because the DM spike would quickly drive an ordinary small eccentricity to a near-extreme value as shown in Fig.~\ref{fig:Res_varybase} with the blue curve drawn from baseline parameters $(e_0=0.1, m_1=10^{10}\,M_\odot, q=1, z=1)$, from which various deviations in the initial eccentricity $e_0$, one of the BH mass $m_1$, the mass ratio $q$, and the redshift $z$ are presented as the orange, purple, green, and red curves, respectively. The efficient growth of orbital eccentricity from a DM spike might as well reproduce the large turn-over eccentricity~\cite{Bi:2023tib} just before the GW-driven stage as inferred from the recent NANOGrav 15-yr data.


\section{Conclusions and discussions}
The DM property in the innermost part of the galactic center is still mysterious to date due to the presence of an SMBH interacting with a DM halo in a complex astrophysical environment during a prolonged history. Such an interplay between SMBH and DM could result in a DM spike even steeper than the cusp, though also possibly flatten if the SMBH grows non-adiabatically. The traditional electromagnetic probe into the DM spike of distant SMBHs might be obscured by the interstellar foreground, and hence the SGWB from SMBHB merger can serve as a clean and transparent probe into the DM spike. 

Recent PTA observations seem to slightly deviate from the simplest explanation from circular inspirals of SMBHBs~\cite{NANOGrav:2023hfp}, which in fact should already depart from the usual single power law $h_c\propto f^{-2/3}$ in the characteristic strain due to the discreteness effects. Nevertheless, even without the motivation from recent PTA observations, calling for more realistic modelings of environmental effects~\cite{Ellis:2023dgf} is still of great theoretical interest for future PTA observations, in particular, from the presence of a DM spike~\cite{Shen:2023pan,Ghoshal:2023fhh}, which should have inevitably introduced the eccentricity for self-consistency in the first place. We therefore in this \textit{paper} carry out a preliminary investigation for the first time into the SGWB from eccentric inspirals of SMBHBs with a DM spike.

We have found a universal infrared scaling in the characteristic strain $h_c\sim f^{7/6-\gamma_\mathrm{sp}/3}$ at low frequencies, and a shallower DM spike of eccentric binaries can also flatten the characteristic strain similar to a steeper DM spike of circular binaries. More intriguingly, even a shallow DM spike can easily drive an ordinarily small initial eccentricity up to the extreme value, featuring an oscillating behavior at low frequencies in addition to the usual turnover-bump-flatter spectrum connecting to the usual single power law at higher frequencies. This not only naturally explains the preferred large turn-over eccentricity seen in the recent PTA data, but also provides a distinctive smoking gun for the DM spike in future PTA observations. Needless to say that such a detection of this oscillation feature around $10^{-10}$Hz is very difficult, requiring successive measurements of Milky-Way pulsars over decades or hundreds of years. Nevertheless, this might as well be feasible from more precise measurements of pulsars outside the Milky Way or the emergence of new detection technologies in future.

Nevertheless, the DM spike is not always subjected to eccentrification but also circularization for, in our case, $\gamma_\mathrm{sp}>3$ as we have checked explicitly, similar to the case of IMRIs as claimed earlier in Refs.~\cite{Yue:2019ozq,Cardoso:2020iji} but revised later in Ref.~\cite{Becker:2021ivq} by considering a velocity distribution for DM particles. Likewise, the preference to circularization was also found to occur for IMRI at $\gamma_\mathrm{sp}>1.5$~\cite{Gould:2003hk} and $\gamma_\mathrm{sp}>1.8$~\cite{Dosopoulou:2023umg} from the standard and improved Chandrasekhar prescriptions of dynamical friction, respectively, by including the contribution from DM particles that move faster than the inspiraling object. This is why we limit ourselves within $\gamma_\mathrm{sp}\leq1.5$ in Fig.~\ref{fig:Res_fixe0} as a steeper DM spike might call for more complex modeling, and our result might as well hint for a flatter DM spike in real world.

Furthermore, the halo feedback effect from the inspiraling object to DM particles should also be included to deplete the local DM halo profile as shown in Refs.~\cite{Kavanagh:2020cfn,Jangra:2023mqp} but only for a circular orbit in the IMRI case, which is, however, very difficult to be precisely modeled for our case of eccentric inspiraling SMBHB, where we simply choose to flatten the DM spike index in Fig.~\ref{fig:Res_fixgamma} down to $\gamma_\mathrm{sp}=0.5$~\cite{Merritt:2002vj} as inferred from considering the energy transfer from SMBHB formation to the DM particles during the halo merger. More realistic modelings including this backreaction effect would necessarily call for numerical relativity simulations (e.g. Ref.~\cite{Yang:2017lpm} for intermediate-mass BH binary merger within an axion-like shell) to settle it down. 

If the DM spike could survive the halo feedback, then the extreme eccentricity driven by a shallow DM spike might instantly lead to head-on collisions of SMBHBs~\cite{Abrahams:1994xy,Anninos:1994vw,Sperhake:2005uf,Witek:2010xi,Witek:2010az,Berti:2010gx,Gwak:2016cbq,Cook:2017fec,Sperhake:2019oaw,CalderonBustillo:2020xms}, causing a super-position of GW bursts~\cite{Fan:2022wio} even before GWs could dominate the evolution. If not, we can put strong constraints on the halo feedback effect from the absence of head-on signals. 

The last concern is how to analytically model these distinctive structures including the oscillation feature we found in the GW spectrum for future data analysis as it is time-consuming to numerically integrate the solved energy spectra over all different population histories. It is also important to give a full survey on the detailed parameter dependence on the SMBH mass-halo mass relation and other SMBHB population histories.

\paragraph{Note added.} This paper~\cite{Alonso-Alvarez:2024gdz} appeared after ours and tackled a very similar system with a similar approach. Their work focuses on a specific type of dark matter and does not take eccentricity into consideration, but does provide a more detailed analysis of PTA data. Our work focus on analyzing the effect of eccentricity when the DM spike is not destroyed (supported by their work~\cite{Alonso-Alvarez:2024gdz} by referring to the self-interacting dark matter).

\acknowledgments
We sincerely thank Niklas Becker, Zhoujian Cao, Siyuan Chen, Xian Chen, Wen-Biao Han, Qing-Guo Huang, Luis C. Ho, Jing Liu, Ning-Qiang Song, Sokratis Trifinopoulos, Wang-Wei Yu, Guan-Wen Yuan, Yefei Yuan, Jian-Dong Zhang, and Yu-Feng Zhou for helpful discussions. 
This work is supported by the National Key Research and Development Program of China Grant No. 2021YFA0718304, No. 2021YFC2203004, and No. 2020YFC2201502,
the National Natural Science Foundation of China Grants No. 12422502,  No. 12105344, No. 12235019, No. 11821505, No. 11991052, and No. 11947302,
the Strategic Priority Research Program of the Chinese Academy of Sciences (CAS) Grant No. XDB23030100, No. XDA15020701, 
the Key Research Program of the CAS Grant No. XDPB15, 
the Key Research Program of Frontier Sciences of CAS,
and the Science Research Grants from the China Manned Space Project with No. CMS-CSST-2021-B01.









\appendix
\section{DM sipke parameters $\rho_\mathrm{sp}$ and $R_\mathrm{sp}$}\label{Spike}


Taking the Navarro-Frenk-White (NFW) profile~\cite{Navarro:1996gj} as an illustrative example,
\begin{align}
\rho_\mathrm{NFW}(r)=\frac{\rho_s}{(r/r_s)(1+r/r_s)^{2}},
\end{align}
the matching condition $\rho_\mathrm{sp}=\rho_\mathrm{spike}(R_\mathrm{sp})=\rho_\mathrm{NFW}(R_\mathrm{sp})$ gives rise to $\rho_\mathrm{sp}$ in terms of $R_\mathrm{sp}$, $\rho_s$, and $r_s$. Note that the spike radius also depends on $\rho_s$ and $r_s$ as well as $M_\mathrm{BH}$ via
\begin{align}
R_\mathrm{sp}=a_\gamma r_s\sqrt{M_\mathrm{BH}/(\rho_sr_s^3)},
\end{align}
with $a_\gamma\simeq0.1$~\cite{Gondolo:1999ef}. To further express $r_s$ and $\rho_s$ in terms of $M_\mathrm{BH}$, we first turn to the concentration $c_{200}(M_{200},z)\equiv R_{200}/r_s$ for the DM halo mass $M_{200}\equiv200\times\frac43\pi\rho_c(z)R_{200}^3$ enclosed within a radius $R_{200}$, where the concentration $c_{200}$ is fitted by Eq.~(24) of Ref.~\cite{Klypin:2014kpa} given $M_{200}$ and $z$, and $R_{200}$ is also determined by $M_{200}$ and $z$ given the critical density 
\begin{align}
\rho_c(z)\equiv\frac{3H_0^2}{8\pi G}\cdot[(1-\Omega_\mathrm{m})+\Omega_\mathrm{m}(1+z)^3],
\end{align}
where $H_0=67.27\,\mathrm{km\, s^{-1}\, Mpc^{-1}}$ and $\Omega_\mathrm{m}=0.3166$~\cite{Planck:2018vyg}. Then, the remaining task is to express $M_{200}$ in terms of $M_\mathrm{BH}$ at $z$ through the bulge mass $M_\mathrm{bulge}$ and stellar mass $M_*$ via~\cite{Kormendy:2013dxa}
\begin{align}
M_\mathrm{BH}=4.9\times10^8\,M_\odot(M_\mathrm{bulge}/10^{11}\,M_\odot)^{1.17},
\end{align}
where  $M_\mathrm{bulge}=0.615M_*$~\cite{Chen:2018znx}, the stellar mass $M_*$ can be used to solve for the halo mass $M_{200}$ via a double power-law function~\cite{Moster:2009fk} fitted by Eqs.~(6-10) with best-fit parameters given in Table 3 of Ref.~\cite{Girelli:2020goz}. Once we obtain $M_{200}$ in terms of $M_\mathrm{BH}$ at $z$, both $c_{200}$ and $R_{200}$ are determined and hence $r_s$ is known correspondingly, then $\rho_s$ can be solved from
\begin{align}
M_{200}=\int_0^{R_{200}}4\pi r^2\rho_\mathrm{NFW}(r)\mathrm{d}r. 
\end{align}

\section{Equations of motion for elliptical orbit}\label{EOM}

The loss rates for the total energy and angular momentum consist of time-averaged contributions from both GW emissions and dynamical frictions as
\begin{align}
\dot{E}
&=\left\langle\frac{\md E}{\md t}\right\rangle_{\rm{GW}}+\left\langle\frac{\md E}{\md t}\right\rangle_{\rm{DF}}=\left\langle\frac{\md E}{\md t}\right\rangle_{\rm{GW}}+\left\langle\frac{\md E_1}{\md t}\right\rangle_{\rm{DF}}+\left\langle\frac{\md E_2}{\md t}\right\rangle_{\rm{DF}},\\
\dot{L}
&=\left\langle\frac{\md L}{\md t}\right\rangle_{\rm{GW}}+\left\langle\frac{\md L}{\md t}\right\rangle_{\rm{DF}}=\left\langle\frac{\md L}{\md t}\right\rangle_{\rm{GW}}+\left\langle\frac{\md L_1}{\md t}\right\rangle_{\rm{DF}}+\left\langle\frac{\md L_2}{\md t}\right\rangle_{\rm{DF}}.
\end{align}
The loss rate caused by GW radiations is already known to the lowest order in post-Newtonian expansion as~\cite{Peters:1964zz}
\begin{align}
\left\langle\frac{\md E}{\md t}\right\rangle_{\rm{GW}}
&=-\frac{32}{5}\frac{G^4m_1^2m_2^2M}{c^5a^5(1-e^2)^{7/2}}\left(1+\frac{73}{24}e^2+\frac{37}{96}e^4\right),\\
\left\langle\frac{\md L}{\md t}\right\rangle_{\rm{GW}}
&=-\frac{32}{5}\frac{G^{7/2}m_1^2m_2^2M^{1/2}}{c^5a^{7/2}(1-e^2)^{2}}\left(1+\frac{7}{8}e^2\right).
\end{align}
For loss rates by dynamical frictions, we need to first derive expressions for $v_1$ and $\dot{\phi}_1$. Note that the energy
\begin{align}
E_1=-\frac{GM_1m_1}{r_1}+\frac{m_1v_1^2}{2}
\end{align}
can be solved for $v_1$ as
\begin{align}
v_1=\sqrt{\frac{2}{m_1}\left(E_1+\frac{GM_1m_1}{r_1}\right)}
=\sqrt{GM_1\left(\frac{2}{r_1}-\frac{1}{a_1}\right)}.
\end{align}
Meanwhile, the angular momentum 
\begin{align}
L_1=m_1r_1^2\dot{\phi}_1 
\end{align}
can also be solved for $\dot{\phi}_1$ as
\begin{align}
\dot{\phi}_1=\frac{L_1}{m_1r_1^2}=\frac{\sqrt{GM_1a_1(1-e^2)}}{r_1^2}.
\end{align}
Then, we can directly calculate the energy loss rate from dynamical frictions as
\begin{align}
\left\langle\frac{\md E_1}{\md t}\right\rangle_\mathrm{DF}
&=\frac{1}{T}\int_{0}^{T}-F_1v_1\md t\nonumber\\
&=-\frac{1}{T}\int_{0}^{T}\md t\frac{4\pi G^2m_1^2\ln\Lambda\rho_\mathrm{sp}R_\mathrm{sp}^{\gamma_\mathrm{sp}}}{r_1^{\gamma_\mathrm{sp}}\sqrt{GM_1\left(\frac{2}{r_1}-\frac{1}{a_1}\right)}}\nonumber\\
&=-\frac{1}{T}\int_{0}^{T}\md t\frac{4\pi G^{3/2}m_1^2\ln\Lambda\rho_\mathrm{sp}R_\mathrm{sp}^{\gamma_\mathrm{sp}}(1+e\cos{\phi_1})^{\gamma_\mathrm{sp}}}{M_1^{1/2}a_1^{\gamma_\mathrm{sp}-1/2}(1-e^2)^{\gamma_\mathrm{sp}-1/2}(1+2e\cos{\phi_1}+e^2)^{1/2}}\nonumber\\
&=-\int_{0}^{2\pi}\md \phi_{1}\frac{2G^{3/2}\ln\Lambda\rho_\mathrm{sp}R_\mathrm{sp}^{\gamma_\mathrm{sp}}(1+e\cos{\phi_1})^{\gamma_\mathrm{sp}-2}}{(1-e^2)^{\gamma_\mathrm{sp}-2}(1+2e\cos{\phi_1}+e^2)^{1/2}}\frac{m_1^2}{M_1^{1/2}a_1^{\gamma_\mathrm{sp}-1/2}}\nonumber\\
&=-\int_{0}^{2\pi}\md \phi_{1}\frac{2G^{3/2}\ln\Lambda\rho_\mathrm{sp}R_\mathrm{sp}^{\gamma_\mathrm{sp}}(1+e\cos{\phi_1})^{\gamma_\mathrm{sp}-2}M^{\gamma_\mathrm{sp}+1/2}}{(1-e^2)^{\gamma_\mathrm{sp}-2}(1+2e\cos{\phi_1}+e^2)^{1/2}a^{\gamma_\mathrm{sp}-1/2}}\frac{m_1^2}{m_2^{\gamma_\mathrm{sp}+1}}.
\end{align}
In the fourth line, we have used the equation $\int_0^T\frac{\md t}{T}(...)=(1-e^2)^{3/2}\int_0^{2\pi}\frac{\md \phi}{2\pi}(1+e\cos{\phi})^{-2}(...)$~\cite{Maggiore:2007ulw}.
Next, we can also directly calculate the loss rate in the angular momentum from dynamical friction by firstly noting that  $\left|\frac{\md L_1}{\md t}\right|=\left|\mathbf{r}_1\times\mathbf{F}_1\right|=r_1F_1\sin\theta$ with $\sin{\theta}=\frac{r_1\dot{\phi_1}}{v_1}$, and hence $\left|\frac{\md L_1}{\md t}\right|=F_1r_1\frac{r_1\dot{\phi_1}}{v_1}$, and then
\begin{align}
\left\langle\frac{dL_1}{\md t}\right\rangle_\mathrm{DF}
&=\frac{1}{T}\int_{0}^{T}-F_1r_1\frac{r_1\dot{\phi_1}}{v_1}dt\nonumber\\
&=-\frac{1}{T}\int_{0}^{T}\md t\frac{4\pi G^2m_1^2\ln\Lambda\rho_\mathrm{sp}R_\mathrm{sp}^{\gamma_\mathrm{sp}}\sqrt{GM_1a_1(1-e^2)}(1+e\cos{\phi_1})^{\gamma_\mathrm{sp}}}{G^{3/2}M_1^{3/2}a_1^{\gamma_\mathrm{sp}-3/2}(1-e^2)^{\gamma_\mathrm{sp}-3/2}(1+2e\cos{\phi_1}+e^2)^{3/2}}\nonumber\\
&=-\int_{0}^{2\pi}\md \phi_{1}\frac{2G\ln\Lambda\rho_\mathrm{sp}R_\mathrm{sp}^{\gamma_\mathrm{sp}}(1+e\cos{\phi_1})^{\gamma_\mathrm{sp}-2}}{(1-e^2)^{\gamma_\mathrm{sp}-7/2}(1+2e\cos{\phi_1}+e^2)^{3/2}}\frac{m_1^2}{M_1a_1^{\gamma_\mathrm{sp}-2}}\nonumber\\
&=-\int_{0}^{2\pi}\md \phi_{1}\frac{2G\ln\Lambda\rho_\mathrm{sp}R_\mathrm{sp}^{\gamma_\mathrm{sp}}(1+e\cos{\phi_1})^{\gamma_\mathrm{sp}-2}M^{\gamma_\mathrm{sp}}}{(1-e^2)^{\gamma_\mathrm{sp}-7/2}(1+2e\cos{\phi_1}+e^2)^{3/2}a^{\gamma_\mathrm{sp}-2}}\frac{m_1^2}{m_2^{\gamma_\mathrm{sp}+1}}.
\end{align}
In the same way, the expressions for $\left\langle\frac{\md E_2}{\md t}\right\rangle_{\rm{DF}}$ and $\left\langle\frac{\md L_2}{\md t}\right\rangle_{\rm{DF}}$ read
\begin{align}
\left\langle\frac{\md E_2}{\md t}\right\rangle_{\rm{DF}}&=-\int_{0}^{2\pi}\md \phi_{2}\frac{2G^{3/2}\ln\Lambda\rho_\mathrm{sp}R_\mathrm{sp}^{\gamma_\mathrm{sp}}(1+e\cos{\phi_2})^{\gamma_\mathrm{sp}-2}M^{\gamma_\mathrm{sp}+1/2}}{(1-e^2)^{\gamma_\mathrm{sp}-2}(1+2e\cos{\phi_2}+e^2)^{1/2}a^{\gamma_\mathrm{sp}-1/2}}\frac{m_2^2}{m_1^{\gamma_\mathrm{sp}+1}},\\
\left\langle\frac{\md L_2}{\md t}\right\rangle_{\rm{DF}}&=-\int_{0}^{2\pi}\md \phi_{2}\frac{2G\ln\Lambda\rho_\mathrm{sp}R_\mathrm{sp}^{\gamma_\mathrm{sp}}(1+e\cos{\phi_2})^{\gamma_\mathrm{sp}-2}M^{\gamma_\mathrm{sp}}}{(1-e^2)^{\gamma_\mathrm{sp}-7/2}(1+2e\cos{\phi_2}+e^2)^{3/2}a^{\gamma_\mathrm{sp}-2}}\frac{m_2^2}{m_1^{\gamma_\mathrm{sp}+1}}.
\end{align}
Finally, the loss rates in the total energy and angular momentum read
\begin{align}\label{Edot}
\dot{E}
&=-\frac{32}{5}\frac{G^4m_1^2m_2^2M}{c^5a^5(1-e^2)^{7/2}}\left(1+\frac{73}{24}e^2+\frac{37}{96}e^4\right)\nonumber\\
&\quad-\int_{0}^{2\pi}\md \phi\frac{2G^{3/2}\ln\Lambda\rho_\mathrm{sp}R_\mathrm{sp}^{\gamma_\mathrm{sp}}(1+e\cos{\phi})^{\gamma_\mathrm{sp}-2}M^{\gamma_\mathrm{sp}+1/2}}{(1-e^2)^{\gamma_\mathrm{sp}-2}(1+2e\cos{\phi}+e^2)^{1/2}a^{\gamma_\mathrm{sp}-1/2}}\left(\frac{m_1^2}{m_2^{\gamma_\mathrm{sp}+1}}+\frac{m_2^2}{m_1^{\gamma_\mathrm{sp}+1}}\right),
\end{align}
\begin{align}\label{Ldot}
\dot{L}
&=-\frac{32}{5}\frac{G^{7/2}m_1^2m_2^2M^{1/2}}{c^5a^{7/2}(1-e^2)^{2}}\left(1+\frac{7}{8}e^2\right)\nonumber\\
&\quad-\int_{0}^{2\pi}\md \phi\frac{2G\ln\Lambda\rho_\mathrm{sp}R_\mathrm{sp}^{\gamma_\mathrm{sp}}(1+e\cos{\phi})^{\gamma_\mathrm{sp}-2}M^{\gamma_\mathrm{sp}}}{(1-e^2)^{\gamma_\mathrm{sp}-7/2}(1+2e\cos{\phi}+e^2)^{3/2}a^{\gamma_\mathrm{sp}-2}}\left(\frac{m_1^2}{m_2^{\gamma_\mathrm{sp}+1}}+\frac{m_2^2}{m_1^{\gamma_\mathrm{sp}+1}}\right).
\end{align}

The dynamical equations~\cite{Peters:1964zz} for eccentric inspirals of SMBHBs with dynamical frictions eventually look like
\begin{align}
\dot{a}&=\frac{2a^2}{Gm_1m_2}\dot{E}\nonumber\\
&=-\frac{64}{5}\frac{G^3m_1m_2M}{c^5a^3(1-e^2)^{7/2}}\left(1+\frac{73}{24}e^2+\frac{37}{96}e^4\right)\nonumber\\
&\quad-\frac{4G^{1/2}\ln\Lambda\rho_\mathrm{sp}R_\mathrm{sp}^{\gamma_\mathrm{sp}}M^{\gamma_\mathrm{sp}+1/2}}{(1-e^2)^{\gamma_\mathrm{sp}-2}a^{\gamma_\mathrm{sp}-5/2}}\left(\frac{m_1}{m_2^{\gamma_\mathrm{sp}+2}}+\frac{m_2}{m_1^{\gamma_\mathrm{sp}+2}}\right)\int_{0}^{2\pi}\md \phi\frac{(1+e\cos{\phi})^{\gamma_\mathrm{sp}-2}}{(1+2e\cos{\phi}+e^2)^{1/2}},\\
\dot{e}&=\frac{a(1-e^2)}{eGm_1m_2}\dot{E}-\frac{\sqrt{M(1-e^2)}}{em_1m_2\sqrt{aG}}\dot{L}\nonumber\\
&=-\frac{304}{15}\frac{G^{3}m_1m_2M}{c^5a^{4}(1-e^2)^{5/2}}e\left(1+\frac{121}{304}e^2\right)\nonumber\\
&\quad-\frac{4G^{1/2}\ln\Lambda\rho_\mathrm{sp}R_\mathrm{sp}^{\gamma_\mathrm{sp}}M^{\gamma_\mathrm{sp}+1/2}}{(1-e^2)^{\gamma_\mathrm{sp}-3}a^{\gamma_\mathrm{sp}-3/2}}\left(\frac{m_1}{m_2^{\gamma_\mathrm{sp}+2}}+\frac{m_2}{m_1^{\gamma_\mathrm{sp}+2}}\right)\int_0^{2\pi}\md \phi\frac{(1+e\cos{\phi})^{\gamma_\mathrm{sp}-2}(e+\cos{\phi})}{(1+2e\cos{\phi}+e^2)^{3/2}}.
\end{align}

\section{The energy density spectrum from the characteristic strain}\label{GW}

The two GW polarizations with true anomaly are shown in Ref.~\cite{Moreno:1995} as
\begin{equation}
\begin{aligned}
h_{+}&=-\frac{G^2M\mu}{c^4a(1-e^2)D_{L}}\bigg[\left(2\cos(2\phi-2\beta)+\frac{5}{2}e\cos(\phi-2\beta)+\frac{1}{2}e\cos(3\phi-2\beta)+e^{2}\cos(2\beta)\right)\\
&\quad\times(1+\cos^{2}\iota)+(e\cos\phi+e^2)\sin^2(\iota)\bigg]\,,\\
h_{\times}&=-{\frac{G^2M\mu}{c^4a(1-e^2)D_{L}}}\left[4\sin(2\phi-2\beta)+5e\sin(\phi-2\beta)+e\sin(3\phi-2\beta)-2e^{2}\sin(2\beta)\right]\cos(\iota),
\end{aligned}
\end{equation}
where $\mu=\frac{m_1m_2}{m_1+m_2}$ is the reduced mass, $D_L$ is the luminosity distance, $\beta$ is the angle between the major axis and the projected direction of the observer onto the orbital plane, and $\iota$ is the angle between the normal direction of the orbital plane and the observer direction. The coefficients of Fourier-Bessel series in the main context are listed below as~\cite{Moore:2018kvz,Becker:2021ivq}
\begin{align}
&C_{+}^{(n)}=\left[2s_{\iota}^{2}J_{n}(n e)+\frac{2}{e^{2}}(1+c_{\iota}^{2})c_{2\beta}\left((e^{2}-2)J_{n}(n e)+n e(1-e^{2})(J_{n-1}(n e)-J_{n+1}(n e))\right)\right],\\
&S_{+}^{(n)}=-\frac{2}{e^{2}}\sqrt{1-e^{2}}(1+c_{\iota}^{2})s_{2\beta}\left[-2(1-e^{2})n J_{n}(n e)+e(J_{n-1}(n e)-J_{n+1}(n e))\right],\\
&C_{\times}^{(n)}=-\frac{4}{e^{2}}c_{\iota}s_{2\beta}\Big[(2-e^{2})J_{n}(n e)+n e(1-e^{2})(J_{n-1}(n e)-J_{n+1}(n e)\Big)\Big],\\
&S_{\times}^{(n)}=-\frac{4}{e^{2}}\sqrt{1-e^{2}}c_{\iota}c_{2\beta}\left[-2(1-e^{2})n J_{n}(n e)+e(J_{n-1}(n e)-J_{n+1}(n e))\right],
\end{align}
where $J_n$ is the Bessel function of the first kind, $c_{\iota}$ and $s_{\iota}$ are the simplified notations for $\cos(\iota)$ and $\sin(\iota)$, so are $c_{\beta}$ and $s_{\beta}$. Therefore, the energy spectrum can be calculated approximately as
\begin{equation}
\begin{aligned}
\frac{\md E_{gw}}{\md f_r}
&=\frac{\pi c^3}{2G}\frac{f^2D_L^2}{(1+z)^2}\int \md \Omega\left(\left|\tilde{h}_{+}(f)\right|^2+\left|\tilde{h}_{\times}(f)\right|^2\right)\\
&=\frac{\pi c^3}{2G}\frac{f^2D_L^2}{(1+z)^2}\int \md \Omega\left(\left|\sum_n\tilde{h}_{+}^{(n)}(f)\right|^2+\left|\sum_n\tilde{h}_{\times}^{(n)}(f)\right|^2\right)\\
&\approx\frac{\pi c^3}{2G}\frac{f^2D_L^2}{(1+z)^2}\int \md \Omega\sum_n\left(\left|\tilde{h}_{+}^{(n)}(f)\right|^2+\left|\tilde{h}_{\times}^{(n)}(f)\right|^2\right),
\end{aligned}
\end{equation}
where the mutually orthogonal approximation~\cite{Moore:2018kvz} between different harmonics are used in the third line. In addition to the characteristic strain $h_c(f)$ in the main text, we also present here shortly below the energy density spectrum
\begin{align}
\Omega_\mathrm{GW}(f)\equiv\frac{1}{\rho_c}\frac{\mathrm{d}\rho_\mathrm{GW}}{\mathrm{d}\ln f}=\frac{\pi}{4}\frac{f^2h_c^2(f)}{G\rho_c}.
\end{align}

\begin{figure}
\centering
\includegraphics[width=0.8\textwidth]{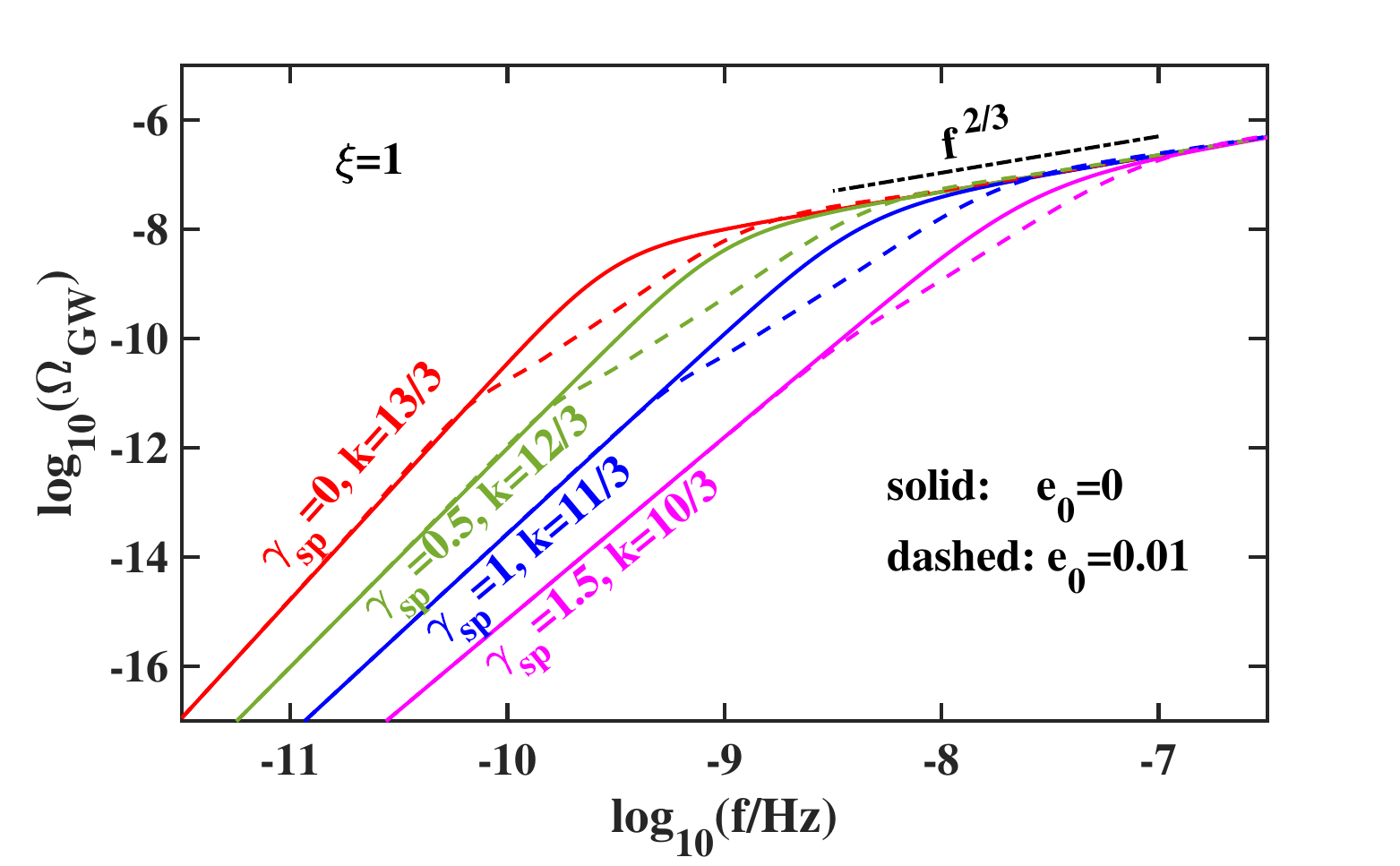}
\caption{The energy density spectrum $\Omega_\mathrm{GW}(f)$ at each GW frequency $f$ for the SMBHB inspiral with negligible initial eccentricity $e_0$ but also with different DM spike index $\gamma_\mathrm{sp}$. $k$ is the fitted infrared slope at low frequencies when DM spike is dominated, and $\xi=1$ is the suppression factor.}
\label{Omega_unfix}
\end{figure}

\begin{figure}
\centering
\includegraphics[width=0.8\textwidth]{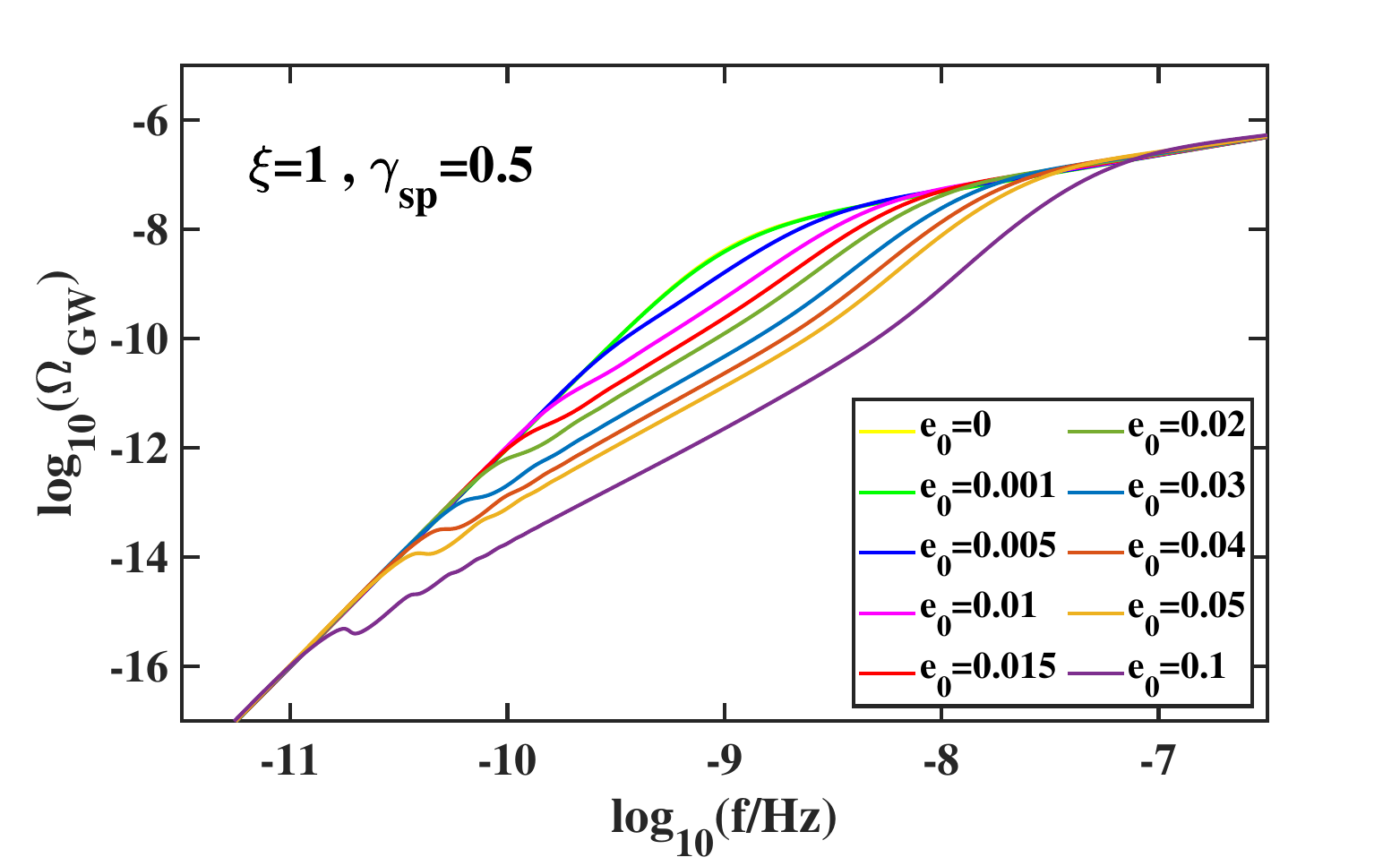}
\caption{The energy density spectrum $\Omega_\mathrm{GW}(f)$ of GWs from SMBHBs with a fixed DM spike index $\gamma_\mathrm{sp}=0.5$, and suppression factor $\xi=1$, but for the initial eccentricity $e_0$ increasing from $0$ to $10^{-1}$.}
\label{Omega_fix}
\end{figure}

\clearpage

\bibliographystyle{JHEP}
\bibliography{ref.bib}

\providecommand{\href}[2]{#2}\begingroup\raggedright\begin{thebibliography}{10}

\bibitem{Rubin:1970zza}
V.C.~Rubin and W.K.~Ford, Jr., \emph{{Rotation of the Andromeda Nebula from a
  Spectroscopic Survey of Emission Regions}},
  \href{https://doi.org/10.1086/150317}{\emph{Astrophys. J.} {\bfseries 159}
  (1970) 379}.

\bibitem{Begeman:1991iy}
K.G.~Begeman, A.H.~Broeils and R.H.~Sanders, \emph{{Extended rotation curves of
  spiral galaxies: Dark haloes and modified dynamics}},
  \href{https://doi.org/10.1093/mnras/249.3.523}{\emph{Mon. Not. Roy. Astron.
  Soc.} {\bfseries 249} (1991) 523}.

\bibitem{Zwicky:1933gu}
F.~Zwicky, \emph{{Die Rotverschiebung von extragalaktischen Nebeln}},
  \href{https://doi.org/10.1007/s10714-008-0707-4}{\emph{Helv. Phys. Acta}
  {\bfseries 6} (1933) 110}.

\bibitem{Clowe:2006eq}
D.~Clowe, M.~Bradac, A.H.~Gonzalez, M.~Markevitch, S.W.~Randall, C.~Jones
  et~al., \emph{{A direct empirical proof of the existence of dark matter}},
  \href{https://doi.org/10.1086/508162}{\emph{Astrophys. J. Lett.} {\bfseries
  648} (2006) L109} [\href{https://arxiv.org/abs/astro-ph/0608407}{{\ttfamily
  astro-ph/0608407}}].

\bibitem{Dietrich:2012mp}
J.P.~Dietrich, N.~Werner, D.~Clowe, A.~Finoguenov, T.~Kitching, L.~Miller
  et~al., \emph{{A filament of dark matter between two clusters of galaxies}},
  \href{https://doi.org/10.1038/nature11224}{\emph{Nature} {\bfseries 487}
  (2012) 202} [\href{https://arxiv.org/abs/1207.0809}{{\ttfamily 1207.0809}}].

\bibitem{Planck:2018vyg}
{\scshape Planck} collaboration, \emph{{Planck 2018 results. VI. Cosmological
  parameters}},
  \href{https://doi.org/10.1051/0004-6361/201833910}{\emph{Astron. Astrophys.}
  {\bfseries 641} (2020) A6}
  [\href{https://arxiv.org/abs/1807.06209}{{\ttfamily 1807.06209}}].

\bibitem{PandaX:2022osq}
{\scshape PandaX} collaboration, \emph{{First Search for the Absorption of
  Fermionic Dark Matter with the PandaX-4T Experiment}},
  \href{https://doi.org/10.1103/PhysRevLett.129.161803}{\emph{Phys. Rev. Lett.}
  {\bfseries 129} (2022) 161803}
  [\href{https://arxiv.org/abs/2205.15771}{{\ttfamily 2205.15771}}].

\bibitem{XENON:2023cxc}
{\scshape XENON} collaboration, \emph{{First Dark Matter Search with Nuclear
  Recoils from the XENONnT Experiment}},
  \href{https://doi.org/10.1103/PhysRevLett.131.041003}{\emph{Phys. Rev. Lett.}
  {\bfseries 131} (2023) 041003}
  [\href{https://arxiv.org/abs/2303.14729}{{\ttfamily 2303.14729}}].

\bibitem{DiMauro:2023qat}
M.~Di~Mauro, J.~P\'erez-Romero, M.A.~S\'anchez-Conde and N.~Fornengo,
  \emph{{Constraining the dark matter contribution of \ensuremath{\gamma} rays
  in clusters of galaxies using Fermi-LAT data}},
  \href{https://doi.org/10.1103/PhysRevD.107.083030}{\emph{Phys. Rev. D}
  {\bfseries 107} (2023) 083030}
  [\href{https://arxiv.org/abs/2303.16930}{{\ttfamily 2303.16930}}].

\bibitem{Baryakhtar:2022hbu}
M.~Baryakhtar et~al., \emph{{Dark Matter In Extreme Astrophysical
  Environments}},  in \emph{{Snowmass 2021}}, 3, 2022
  [\href{https://arxiv.org/abs/2203.07984}{{\ttfamily 2203.07984}}].

\bibitem{Richstone:1998ky}
D.~Richstone et~al., \emph{{Supermassive black holes and the evolution of
  galaxies}}, {\emph{Nature} {\bfseries 395} (1998) A14}
  [\href{https://arxiv.org/abs/astro-ph/9810378}{{\ttfamily
  astro-ph/9810378}}].

\bibitem{Kormendy:2013dxa}
J.~Kormendy and L.C.~Ho, \emph{{Coevolution (Or Not) of Supermassive Black
  Holes and Host Galaxies}},
  \href{https://doi.org/10.1146/annurev-astro-082708-101811}{\emph{Ann. Rev.
  Astron. Astrophys.} {\bfseries 51} (2013) 511}
  [\href{https://arxiv.org/abs/1304.7762}{{\ttfamily 1304.7762}}].

\bibitem{Gondolo:1999ef}
P.~Gondolo and J.~Silk, \emph{{Dark matter annihilation at the galactic
  center}}, \href{https://doi.org/10.1103/PhysRevLett.83.1719}{\emph{Phys. Rev.
  Lett.} {\bfseries 83} (1999) 1719}
  [\href{https://arxiv.org/abs/astro-ph/9906391}{{\ttfamily
  astro-ph/9906391}}].

\bibitem{Ferrer:2017xwm}
F.~Ferrer, A.M.~da~Rosa and C.M.~Will, \emph{{Dark matter spikes in the
  vicinity of Kerr black holes}},
  \href{https://doi.org/10.1103/PhysRevD.96.083014}{\emph{Phys. Rev. D}
  {\bfseries 96} (2017) 083014}
  [\href{https://arxiv.org/abs/1707.06302}{{\ttfamily 1707.06302}}].

\bibitem{Lotz:2011cn}
J.M.~Lotz, P.~Jonsson, T.J.~Cox, D.~Croton, J.R.~Primack, R.S.~Somerville
  et~al., \emph{{The Major and Minor Galaxy Merger Rates at z \ensuremath{<}
  1.5}}, \href{https://doi.org/10.1088/0004-637X/742/2/103}{\emph{Astrophys.
  J.} {\bfseries 742} (2011) 103}
  [\href{https://arxiv.org/abs/1108.2508}{{\ttfamily 1108.2508}}].

\bibitem{Lacey:1993iv}
C.G.~Lacey and S.~Cole, \emph{{Merger rates in hierarchical models of galaxy
  formation}}, {\emph{Mon. Not. Roy. Astron. Soc.} {\bfseries 262} (1993) 627}.

\bibitem{DeRosa:2019myq}
A.~De~Rosa et~al., \emph{{The quest for dual and binary supermassive black
  holes: A multi-messenger view}},
  \href{https://doi.org/10.1016/j.newar.2020.101525}{\emph{New Astron. Rev.}
  {\bfseries 86} (2019) 101525}
  [\href{https://arxiv.org/abs/2001.06293}{{\ttfamily 2001.06293}}].

\bibitem{Barnes:1992rm}
J.E.~Barnes and L.E.~Hernquist, \emph{{Dynamics of interacting galaxies}},
  \href{https://doi.org/10.1146/annurev.aa.30.090192.003421}{\emph{Ann. Rev.
  Astron. Astrophys.} {\bfseries 30} (1992) 705}.

\bibitem{Chandrasekhar:1943ys}
S.~Chandrasekhar, \emph{{Dynamical Friction. I. General Considerations: the
  Coefficient of Dynamical Friction}},
  \href{https://doi.org/10.1086/144517}{\emph{Astrophys. J.} {\bfseries 97}
  (1943) 255}.

\bibitem{Antonini:2011tu}
F.~Antonini and D.~Merritt, \emph{{Dynamical Friction around Supermassive Black
  Holes}}, \href{https://doi.org/10.1088/0004-637X/745/1/83}{\emph{Astrophys.
  J.} {\bfseries 745} (2012) 83}
  [\href{https://arxiv.org/abs/1108.1163}{{\ttfamily 1108.1163}}].

\bibitem{Begelman:1980vb}
M.C.~Begelman, R.D.~Blandford and M.J.~Rees, \emph{{Massive black hole binaries
  in active galactic nuclei}},
  \href{https://doi.org/10.1038/287307a0}{\emph{Nature} {\bfseries 287} (1980)
  307}.

\bibitem{Merritt:2004gc}
D.~Merritt and M.~Milosavljevic, \emph{{Massive black hole binary evolution}},
  \href{https://doi.org/10.12942/lrr-2005-8}{\emph{Living Rev. Rel.} {\bfseries
  8} (2005) 8} [\href{https://arxiv.org/abs/astro-ph/0410364}{{\ttfamily
  astro-ph/0410364}}].

\bibitem{Peters:1964zz}
P.C.~Peters, \emph{{Gravitational Radiation and the Motion of Two Point
  Masses}}, \href{https://doi.org/10.1103/PhysRev.136.B1224}{\emph{Phys. Rev.}
  {\bfseries 136} (1964) B1224}.

\bibitem{Yu:2001xp}
Q.~Yu, \emph{{Evolution of massive binary black holes}},
  \href{https://doi.org/10.1046/j.1365-8711.2002.05242.x}{\emph{Mon. Not. Roy.
  Astron. Soc.} {\bfseries 331} (2002) 935}
  [\href{https://arxiv.org/abs/astro-ph/0109530}{{\ttfamily
  astro-ph/0109530}}].

\bibitem{Sesana:2013wja}
A.~Sesana, \emph{{Insights into the astrophysics of supermassive black hole
  binaries from pulsar timing observations}},
  \href{https://doi.org/10.1088/0264-9381/30/22/224014}{\emph{Class. Quant.
  Grav.} {\bfseries 30} (2013) 224014}
  [\href{https://arxiv.org/abs/1307.2600}{{\ttfamily 1307.2600}}].

\bibitem{2012AdAst2012E...3D}
M.~{Dotti}, A.~{Sesana} and R.~{Decarli}, \emph{{Massive Black Hole Binaries:
  Dynamical Evolution and Observational Signatures}},
  \href{https://doi.org/10.1155/2012/940568}{\emph{Advances in Astronomy}
  {\bfseries 2012} (2012) 940568}
  [\href{https://arxiv.org/abs/1111.0664}{{\ttfamily 1111.0664}}].

\bibitem{Alonso-Alvarez:2024gdz}
G.~Alonso-\'Alvarez, J.M.~Cline and C.~Dewar, \emph{{Self-Interacting Dark
  Matter Solves the Final Parsec Problem of Supermassive Black Hole Mergers}},
  \href{https://doi.org/10.1103/PhysRevLett.133.021401}{\emph{Phys. Rev. Lett.}
  {\bfseries 133} (2024) 021401}
  [\href{https://arxiv.org/abs/2401.14450}{{\ttfamily 2401.14450}}].

\bibitem{NANOGrav:2019tvo}
{\scshape NANOGrav} collaboration, \emph{{Multi-Messenger Astrophysics with
  Pulsar Timing Arrays}},  \href{https://arxiv.org/abs/1903.07644}{{\ttfamily
  1903.07644}}.

\bibitem{Bi:2023tib}
Y.-C.~Bi, Y.-M.~Wu, Z.-C.~Chen and Q.-G.~Huang, \emph{{Implications for the
  Supermassive Black Hole Binaries from the NANOGrav 15-year Data Set}},
  \href{https://doi.org/10.1007/s11433-023-2252-4}{\emph{Sci. China Phys. Mech.
  Astron.} {\bfseries 66} (2023) 120402}
  [\href{https://arxiv.org/abs/2307.00722}{{\ttfamily 2307.00722}}].

\bibitem{Daghigh:2022pcr}
R.G.~Daghigh and G.~Kunstatter, \emph{{Spacetime Metrics and Ringdown Waveforms
  for Galactic Black Holes Surrounded by a Dark Matter Spike}},
  \href{https://doi.org/10.3847/1538-4357/ac940b}{\emph{Astrophys. J.}
  {\bfseries 940} (2022) 33}
  [\href{https://arxiv.org/abs/2206.04195}{{\ttfamily 2206.04195}}].

\bibitem{NANOGrav:2023gor}
{\scshape NANOGrav} collaboration, \emph{{The NANOGrav 15 yr Data Set: Evidence
  for a Gravitational-wave Background}},
  \href{https://doi.org/10.3847/2041-8213/acdac6}{\emph{Astrophys. J. Lett.}
  {\bfseries 951} (2023) L8}
  [\href{https://arxiv.org/abs/2306.16213}{{\ttfamily 2306.16213}}].

\bibitem{EPTA:2023sfo}
{\scshape EPTA} collaboration, \emph{{The second data release from the European
  Pulsar Timing Array - I. The dataset and timing analysis}},
  \href{https://doi.org/10.1051/0004-6361/202346841}{\emph{Astron. Astrophys.}
  {\bfseries 678} (2023) A48}
  [\href{https://arxiv.org/abs/2306.16224}{{\ttfamily 2306.16224}}].

\bibitem{Reardon:2023gzh}
D.J.~Reardon et~al., \emph{{Search for an Isotropic Gravitational-wave
  Background with the Parkes Pulsar Timing Array}},
  \href{https://doi.org/10.3847/2041-8213/acdd02}{\emph{Astrophys. J. Lett.}
  {\bfseries 951} (2023) L6}
  [\href{https://arxiv.org/abs/2306.16215}{{\ttfamily 2306.16215}}].

\bibitem{Xu:2023wog}
H.~Xu et~al., \emph{{Searching for the Nano-Hertz Stochastic Gravitational Wave
  Background with the Chinese Pulsar Timing Array Data Release I}},
  \href{https://doi.org/10.1088/1674-4527/acdfa5}{\emph{Res. Astron.
  Astrophys.} {\bfseries 23} (2023) 075024}
  [\href{https://arxiv.org/abs/2306.16216}{{\ttfamily 2306.16216}}].

\bibitem{NANOGrav:2023hfp}
{\scshape NANOGrav} collaboration, \emph{{The NANOGrav 15 yr Data Set:
  Constraints on Supermassive Black Hole Binaries from the Gravitational-wave
  Background}},
  \href{https://doi.org/10.3847/2041-8213/ace18b}{\emph{Astrophys. J. Lett.}
  {\bfseries 952} (2023) L37}
  [\href{https://arxiv.org/abs/2306.16220}{{\ttfamily 2306.16220}}].

\bibitem{Maggiore:2018sht}
M.~Maggiore, \emph{{Gravitational Waves. Vol. 2: Astrophysics and Cosmology}},
  Oxford University Press (3, 2018).

\bibitem{Ellis:2023dgf}
J.~Ellis, M.~Fairbairn, G.~H\"utsi, J.~Raidal, J.~Urrutia, V.~Vaskonen et~al.,
  \emph{{Gravitational Waves from SMBH Binaries in Light of the NANOGrav
  15-Year Data}},  \href{https://arxiv.org/abs/2306.17021}{{\ttfamily
  2306.17021}}.

\bibitem{Ghoshal:2023fhh}
A.~Ghoshal and A.~Strumia, \emph{{Probing the Dark Matter density with
  gravitational waves from super-massive binary black holes}},
  \href{https://arxiv.org/abs/2306.17158}{{\ttfamily 2306.17158}}.

\bibitem{Shen:2023pan}
Z.-Q.~Shen, G.-W.~Yuan, Y.-Y.~Wang and Y.-Z.~Wang, \emph{{Dark Matter Spike
  surrounding Supermassive Black Holes Binary and the nanohertz Stochastic
  Gravitational Wave Background}},
  \href{https://arxiv.org/abs/2306.17143}{{\ttfamily 2306.17143}}.

\bibitem{Aghaie:2023lan}
M.~Aghaie, G.~Armando, A.~Dondarini and P.~Panci, \emph{{Bounds on Ultralight
  Dark Matter from NANOGrav}},
  \href{https://arxiv.org/abs/2308.04590}{{\ttfamily 2308.04590}}.

\bibitem{EPTA:2023xxk}
{\scshape EPTA, InPTA} collaboration, \emph{{The second data release from the
  European Pulsar Timing Array - IV. Implications for massive black holes, dark
  matter, and the early Universe}},
  \href{https://doi.org/10.1051/0004-6361/202347433}{\emph{Astron. Astrophys.}
  {\bfseries 685} (2024) A94}
  [\href{https://arxiv.org/abs/2306.16227}{{\ttfamily 2306.16227}}].

\bibitem{Agazie:2024jbf}
G.~Agazie et~al., \emph{{The NANOGrav 15 yr Data Set: Looking for Signs of
  Discreteness in the Gravitational-wave Background}},
  \href{https://arxiv.org/abs/2404.07020}{{\ttfamily 2404.07020}}.

\bibitem{Yue:2019ozq}
X.-J.~Yue and Z.~Cao, \emph{{Dark matter minispike: A significant enhancement
  of eccentricity for intermediate-mass-ratio inspirals}},
  \href{https://doi.org/10.1103/PhysRevD.100.043013}{\emph{Phys. Rev. D}
  {\bfseries 100} (2019) 043013}
  [\href{https://arxiv.org/abs/1908.10241}{{\ttfamily 1908.10241}}].

\bibitem{Enoki:2006kj}
M.~Enoki and M.~Nagashima, \emph{{The Effect of Orbital Eccentricity on
  Gravitational Wave Background Radiation from Cosmological Binaries}},
  \href{https://doi.org/10.1143/PTP.117.241}{\emph{Prog. Theor. Phys.}
  {\bfseries 117} (2007) 241}
  [\href{https://arxiv.org/abs/astro-ph/0609377}{{\ttfamily
  astro-ph/0609377}}].

\bibitem{Huerta:2015pva}
E.A.~Huerta, S.T.~McWilliams, J.R.~Gair and S.R.~Taylor, \emph{{Detection of
  eccentric supermassive black hole binaries with pulsar timing arrays:
  Signal-to-noise ratio calculations}},
  \href{https://doi.org/10.1103/PhysRevD.92.063010}{\emph{Phys. Rev. D}
  {\bfseries 92} (2015) 063010}
  [\href{https://arxiv.org/abs/1504.00928}{{\ttfamily 1504.00928}}].

\bibitem{Chen:2016zyo}
S.~Chen, A.~Sesana and W.~Del~Pozzo, \emph{{Efficient computation of the
  gravitational wave spectrum emitted by eccentric massive black hole binaries
  in stellar environments}},
  \href{https://doi.org/10.1093/mnras/stx1093}{\emph{Mon. Not. Roy. Astron.
  Soc.} {\bfseries 470} (2017) 1738}
  [\href{https://arxiv.org/abs/1612.00455}{{\ttfamily 1612.00455}}].

\bibitem{Kelley:2017lek}
L.Z.~Kelley, L.~Blecha, L.~Hernquist, A.~Sesana and S.R.~Taylor, \emph{{The
  Gravitational Wave Background from Massive Black Hole Binaries in Illustris:
  spectral features and time to detection with pulsar timing arrays}},
  \href{https://doi.org/10.1093/mnras/stx1638}{\emph{Mon. Not. Roy. Astron.
  Soc.} {\bfseries 471} (2017) 4508}
  [\href{https://arxiv.org/abs/1702.02180}{{\ttfamily 1702.02180}}].

\bibitem{Ostriker:1999ee}
J.P.~Ostriker, \emph{{Collisional dark matter and the origin of massive black
  holes}}, \href{https://doi.org/10.1103/PhysRevLett.84.5258}{\emph{Phys. Rev.
  Lett.} {\bfseries 84} (2000) 5258}
  [\href{https://arxiv.org/abs/astro-ph/9912548}{{\ttfamily
  astro-ph/9912548}}].

\bibitem{Ullio:2001fb}
P.~Ullio, H.~Zhao and M.~Kamionkowski, \emph{{A Dark matter spike at the
  galactic center?}},
  \href{https://doi.org/10.1103/PhysRevD.64.043504}{\emph{Phys. Rev. D}
  {\bfseries 64} (2001) 043504}
  [\href{https://arxiv.org/abs/astro-ph/0101481}{{\ttfamily
  astro-ph/0101481}}].

\bibitem{Merritt:2002vj}
D.~Merritt, M.~Milosavljevic, L.~Verde and R.~Jimenez, \emph{{Dark matter
  spikes and annihilation radiation from the galactic center}},
  \href{https://doi.org/10.1103/PhysRevLett.88.191301}{\emph{Phys. Rev. Lett.}
  {\bfseries 88} (2002) 191301}
  [\href{https://arxiv.org/abs/astro-ph/0201376}{{\ttfamily
  astro-ph/0201376}}].

\bibitem{Gnedin:2003rj}
O.Y.~Gnedin and J.R.~Primack, \emph{{Dark Matter Profile in the Galactic
  Center}}, \href{https://doi.org/10.1103/PhysRevLett.93.061302}{\emph{Phys.
  Rev. Lett.} {\bfseries 93} (2004) 061302}
  [\href{https://arxiv.org/abs/astro-ph/0308385}{{\ttfamily
  astro-ph/0308385}}].

\bibitem{Merritt:2003pf}
D.~Merritt and M.Y.~Poon, \emph{{Chaotic loss cones, black hole fueling and the
  m-sigma relation}}, \href{https://doi.org/10.1086/382497}{\emph{Astrophys.
  J.} {\bfseries 606} (2004) 788}
  [\href{https://arxiv.org/abs/astro-ph/0302296}{{\ttfamily
  astro-ph/0302296}}].

\bibitem{Merritt:2003qk}
D.~Merritt, \emph{{Evolution of the dark matter distribution at the galactic
  center}}, \href{https://doi.org/10.1103/PhysRevLett.92.201304}{\emph{Phys.
  Rev. Lett.} {\bfseries 92} (2004) 201304}
  [\href{https://arxiv.org/abs/astro-ph/0311594}{{\ttfamily
  astro-ph/0311594}}].

\bibitem{Shapiro:2022prq}
S.L.~Shapiro and D.C.~Heggie, \emph{{Effect of stars on the dark matter spike
  around a black hole: A tale of two treatments}},
  \href{https://doi.org/10.1103/PhysRevD.106.043018}{\emph{Phys. Rev. D}
  {\bfseries 106} (2022) 043018}
  [\href{https://arxiv.org/abs/2209.08105}{{\ttfamily 2209.08105}}].

\bibitem{Navarro:1996gj}
J.F.~Navarro, C.S.~Frenk and S.D.M.~White, \emph{{A Universal density profile
  from hierarchical clustering}},
  \href{https://doi.org/10.1086/304888}{\emph{Astrophys. J.} {\bfseries 490}
  (1997) 493} [\href{https://arxiv.org/abs/astro-ph/9611107}{{\ttfamily
  astro-ph/9611107}}].

\bibitem{Binney:2008}
J.~{Binney} and S.~{Tremaine}, \emph{Galactic Dynamics: Second Edition},
  Princeton University Press (2008).

\bibitem{Dosopoulou:2023umg}
F.~Dosopoulou, \emph{{Dynamical friction in dark matter spikes: corrections to
  Chandrasekhar's formula}},
  \href{https://arxiv.org/abs/2305.17281}{{\ttfamily 2305.17281}}.

\bibitem{Burke-Spolaor:2018bvk}
S.~Burke-Spolaor et~al., \emph{{The Astrophysics of Nanohertz Gravitational
  Waves}}, \href{https://doi.org/10.1007/s00159-019-0115-7}{\emph{Astron.
  Astrophys. Rev.} {\bfseries 27} (2019) 5}
  [\href{https://arxiv.org/abs/1811.08826}{{\ttfamily 1811.08826}}].

\bibitem{Moore:2018kvz}
B.~Moore, T.~Robson, N.~Loutrel and N.~Yunes, \emph{{Towards a Fourier domain
  waveform for non-spinning binaries with arbitrary eccentricity}},
  \href{https://doi.org/10.1088/1361-6382/aaea00}{\emph{Class. Quant. Grav.}
  {\bfseries 35} (2018) 235006}
  [\href{https://arxiv.org/abs/1807.07163}{{\ttfamily 1807.07163}}].

\bibitem{Becker:2021ivq}
N.~Becker, L.~Sagunski, L.~Prinz and S.~Rastgoo, \emph{{Circularization versus
  eccentrification in intermediate mass ratio inspirals inside dark matter
  spikes}}, \href{https://doi.org/10.1103/PhysRevD.105.063029}{\emph{Phys. Rev.
  D} {\bfseries 105} (2022) 063029}
  [\href{https://arxiv.org/abs/2112.09586}{{\ttfamily 2112.09586}}].

\bibitem{Maggiore:2007ulw}
M.~Maggiore, \emph{{Gravitational Waves. Vol. 1: Theory and Experiments}},
  Oxford University Press (2007),
  \href{https://doi.org/10.1093/acprof:oso/9780198570745.001.0001}{10.1093/acprof:oso/9780198570745.001.0001}.

\bibitem{Chen:2018znx}
S.~Chen, A.~Sesana and C.J.~Conselice, \emph{{Constraining astrophysical
  observables of Galaxy and Supermassive Black Hole Binary Mergers using Pulsar
  Timing Arrays}}, \href{https://doi.org/10.1093/mnras/stz1722}{\emph{Mon. Not.
  Roy. Astron. Soc.} {\bfseries 488} (2019) 401}
  [\href{https://arxiv.org/abs/1810.04184}{{\ttfamily 1810.04184}}].

\bibitem{Shen:2023kkm}
Z.-Q.~Shen, G.-W.~Yuan, C.-Z.~Jiang, Y.-L.S.~Tsai, Q.~Yuan and Y.-Z.~Fan,
  \emph{{Exploring dark matter spike distribution around the Galactic centre
  with stellar orbits}},
  \href{https://doi.org/10.1093/mnras/stad3282}{\emph{Mon. Not. Roy. Astron.
  Soc.} {\bfseries 527} (2024) 3196}
  [\href{https://arxiv.org/abs/2303.09284}{{\ttfamily 2303.09284}}].

\bibitem{Cardoso:2020iji}
V.~Cardoso, C.F.B.~Macedo and R.~Vicente, \emph{{Eccentricity evolution of
  compact binaries and applications to gravitational-wave physics}},
  \href{https://doi.org/10.1103/PhysRevD.103.023015}{\emph{Phys. Rev. D}
  {\bfseries 103} (2021) 023015}
  [\href{https://arxiv.org/abs/2010.15151}{{\ttfamily 2010.15151}}].

\bibitem{Gould:2003hk}
A.~Gould and A.C.~Quillen, \emph{{Sgr A* companion S0-2: A probe of very
  high-mass star formation}},
  \href{https://doi.org/10.1086/375840}{\emph{Astrophys. J.} {\bfseries 592}
  (2003) 935} [\href{https://arxiv.org/abs/astro-ph/0302437}{{\ttfamily
  astro-ph/0302437}}].

\bibitem{Kavanagh:2020cfn}
B.J.~Kavanagh, D.A.~Nichols, G.~Bertone and D.~Gaggero, \emph{{Detecting dark
  matter around black holes with gravitational waves: Effects of dark-matter
  dynamics on the gravitational waveform}},
  \href{https://doi.org/10.1103/PhysRevD.102.083006}{\emph{Phys. Rev. D}
  {\bfseries 102} (2020) 083006}
  [\href{https://arxiv.org/abs/2002.12811}{{\ttfamily 2002.12811}}].

\bibitem{Jangra:2023mqp}
P.~Jangra, B.J.~Kavanagh and J.M.~Diego, \emph{{Impact of dark matter spikes on
  the merger rates of Primordial Black Holes}},
  \href{https://doi.org/10.1088/1475-7516/2023/11/069}{\emph{JCAP} {\bfseries
  11} (2023) 069} [\href{https://arxiv.org/abs/2304.05892}{{\ttfamily
  2304.05892}}].

\bibitem{Yang:2017lpm}
Q.~Yang, L.-W.~Ji, B.~Hu, Z.-J.~Cao and R.-G.~Cai, \emph{{An axion-like scalar
  field environment effect on binary black hole merger}},
  \href{https://doi.org/10.1088/1674-4527/18/6/65}{\emph{Res. Astron.
  Astrophys.} {\bfseries 18} (2018) 065}
  [\href{https://arxiv.org/abs/1706.00678}{{\ttfamily 1706.00678}}].

\bibitem{Abrahams:1994xy}
A.M.~Abrahams, S.L.~Shapiro and S.A.~Teukolsky, \emph{{Calculation of
  gravitational wave forms from black hole collisions and disk collapse:
  Applying perturbation theory to numerical space-times}},
  \href{https://doi.org/10.1103/PhysRevD.51.4295}{\emph{Phys. Rev. D}
  {\bfseries 51} (1995) 4295}
  [\href{https://arxiv.org/abs/gr-qc/9408036}{{\ttfamily gr-qc/9408036}}].

\bibitem{Anninos:1994vw}
P.~Anninos, D.~Hobill, E.~Seidel, L.~Smarr and W.-M.~Suen, \emph{{The Headon
  collision of two equal mass black holes: Numerical methods}},
  \href{https://arxiv.org/abs/gr-qc/9408042}{{\ttfamily gr-qc/9408042}}.

\bibitem{Sperhake:2005uf}
U.~Sperhake, B.J.~Kelly, P.~Laguna, K.L.~Smith and E.~Schnetter, \emph{{Black
  hole head-on collisions and gravitational waves with fixed mesh-refinement
  and dynamic singularity excision}},
  \href{https://doi.org/10.1103/PhysRevD.71.124042}{\emph{Phys. Rev. D}
  {\bfseries 71} (2005) 124042}
  [\href{https://arxiv.org/abs/gr-qc/0503071}{{\ttfamily gr-qc/0503071}}].

\bibitem{Witek:2010xi}
H.~Witek, M.~Zilhao, L.~Gualtieri, V.~Cardoso, C.~Herdeiro, A.~Nerozzi et~al.,
  \emph{{Numerical relativity for D dimensional space-times: head-on collisions
  of black holes and gravitational wave extraction}},
  \href{https://doi.org/10.1103/PhysRevD.82.104014}{\emph{Phys. Rev. D}
  {\bfseries 82} (2010) 104014}
  [\href{https://arxiv.org/abs/1006.3081}{{\ttfamily 1006.3081}}].

\bibitem{Witek:2010az}
H.~Witek, V.~Cardoso, L.~Gualtieri, C.~Herdeiro, U.~Sperhake and M.~Zilhao,
  \emph{{Head-on collisions of unequal mass black holes in D=5 dimensions}},
  \href{https://doi.org/10.1103/PhysRevD.83.044017}{\emph{Phys. Rev. D}
  {\bfseries 83} (2011) 044017}
  [\href{https://arxiv.org/abs/1011.0742}{{\ttfamily 1011.0742}}].

\bibitem{Berti:2010gx}
E.~Berti, V.~Cardoso and B.~Kipapa, \emph{{Up to eleven: radiation from
  particles with arbitrary energy falling into higher-dimensional black
  holes}}, \href{https://doi.org/10.1103/PhysRevD.83.084018}{\emph{Phys. Rev.
  D} {\bfseries 83} (2011) 084018}
  [\href{https://arxiv.org/abs/1010.3874}{{\ttfamily 1010.3874}}].

\bibitem{Gwak:2016cbq}
B.~Gwak and B.-H.~Lee, \emph{{The Upper Bound of Radiation Energy in the
  Myers-Perry Black Hole Collision}},
  \href{https://doi.org/10.1007/JHEP07(2016)079}{\emph{JHEP} {\bfseries 07}
  (2016) 079} [\href{https://arxiv.org/abs/1603.02103}{{\ttfamily
  1603.02103}}].

\bibitem{Cook:2017fec}
W.G.~Cook, U.~Sperhake, E.~Berti and V.~Cardoso, \emph{{Black-hole head-on
  collisions in higher dimensions}},
  \href{https://doi.org/10.1103/PhysRevD.96.124006}{\emph{Phys. Rev. D}
  {\bfseries 96} (2017) 124006}
  [\href{https://arxiv.org/abs/1709.10514}{{\ttfamily 1709.10514}}].

\bibitem{Sperhake:2019oaw}
U.~Sperhake, W.~Cook and D.~Wang, \emph{{High-energy collision of black holes
  in higher dimensions}},
  \href{https://doi.org/10.1103/PhysRevD.100.104046}{\emph{Phys. Rev. D}
  {\bfseries 100} (2019) 104046}
  [\href{https://arxiv.org/abs/1909.02997}{{\ttfamily 1909.02997}}].

\bibitem{CalderonBustillo:2020xms}
J.~Calder\'on~Bustillo, N.~Sanchis-Gual, A.~Torres-Forn\'e and J.A.~Font,
  \emph{{Confusing Head-On Collisions with Precessing Intermediate-Mass Binary
  Black Hole Mergers}},
  \href{https://doi.org/10.1103/PhysRevLett.126.201101}{\emph{Phys. Rev. Lett.}
  {\bfseries 126} (2021) 201101}
  [\href{https://arxiv.org/abs/2009.01066}{{\ttfamily 2009.01066}}].

\bibitem{Fan:2022wio}
H.-M.~Fan, S.~Zhong, Z.-C.~Liang, Z.~Wu, J.-d.~Zhang and Y.-M.~Hu,
  \emph{{Extreme-mass-ratio burst detection with TianQin}},
  \href{https://doi.org/10.1103/PhysRevD.106.124028}{\emph{Phys. Rev. D}
  {\bfseries 106} (2022) 124028}
  [\href{https://arxiv.org/abs/2209.13387}{{\ttfamily 2209.13387}}].

\bibitem{Klypin:2014kpa}
A.~Klypin, G.~Yepes, S.~Gottlober, F.~Prada and S.~Hess, \emph{{MultiDark
  simulations: the story of dark matter halo concentrations and density
  profiles}}, \href{https://doi.org/10.1093/mnras/stw248}{\emph{Mon. Not. Roy.
  Astron. Soc.} {\bfseries 457} (2016) 4340}
  [\href{https://arxiv.org/abs/1411.4001}{{\ttfamily 1411.4001}}].

\bibitem{Moster:2009fk}
B.P.~Moster, R.S.~Somerville, C.~Maulbetsch, F.C.v.d.~Bosch, A.V.~Maccio',
  T.~Naab et~al., \emph{{Constraints on the relationship between stellar mass
  and halo mass at low and high redshift}},
  \href{https://doi.org/10.1088/0004-637X/710/2/903}{\emph{Astrophys. J.}
  {\bfseries 710} (2010) 903}
  [\href{https://arxiv.org/abs/0903.4682}{{\ttfamily 0903.4682}}].

\bibitem{Girelli:2020goz}
G.~Girelli, L.~Pozzetti, M.~Bolzonella, C.~Giocoli, F.~Marulli and M.~Baldi,
  \emph{{The stellar-to-halo mass relation over the past 12 Gyr: I. Standard
  $\Lambda$CDM model}},
  \href{https://doi.org/10.1051/0004-6361/201936329}{\emph{Astron. Astrophys.}
  {\bfseries 634} (2020) A135}
  [\href{https://arxiv.org/abs/2001.02230}{{\ttfamily 2001.02230}}].

\bibitem{Moreno:1995}
C.~Moreno-Garrido, E.~Mediavilla and J.~Buitrago, \emph{{Gravitational
  radiation from point masses in elliptical orbits: spectral analysis and
  orbital parameters}},
  \href{https://doi.org/10.1093/mnras/274.1.115}{\emph{Monthly Notices of the
  Royal Astronomical Society} {\bfseries 274} (1995) 115}
  [\href{https://arxiv.org/abs/https://academic.oup.com/mnras/article-pdf/274/1/115/18539844/mnras274-0115.pdf}{{\ttfamily
  https://academic.oup.com/mnras/article-pdf/274/1/115/18539844/mnras274-0115.pdf}}].

\end{thebibliography}\endgroup


\end{document}